\documentclass[%
 aip,
 amsmath,amssymb,
 reprint,%
]{revtex4-1}

\usepackage{graphicx}
\usepackage{dcolumn}
\usepackage{bm}
\usepackage{xcolor}
\usepackage[utf8]{inputenc}
\usepackage[T1]{fontenc}
\usepackage{mathptmx}
\usepackage{etoolbox}
\usepackage{float}

\makeatletter
\def\@email#1#2{%
 \endgroup
 \patchcmd{\titleblock@produce}
  {\frontmatter@RRAPformat}
  {\frontmatter@RRAPformat{\produce@RRAP{*#1\href{mailto:#2}{#2}}}\frontmatter@RRAPformat}
  {}{}
}%
\makeatother
\begin{document}
\preprint{AIP/123-QED}

\title{Local flow control by phononic subsurfaces over extended spatial domains}
\author{Armin Kianfar}
\affiliation{Ann and H.J. Smead Department of Aerospace Engineering Sciences, University of Colorado Boulder, Boulder, Colorado 80303, USA}
\author{Mahmoud I. Hussein}%
\email{mih@colorado.edu}
\affiliation{Ann and H.J. Smead Department of Aerospace Engineering Sciences, University of Colorado Boulder, Boulder, Colorado 80303, USA}
\affiliation{Department of Physics, University of Colorado Boulder, Boulder, Colorado 80302, USA}


\begin{abstract}
Local phonon motion underneath a surface interacting with a flow may cause the flow to passively stabilize, or destabilize, as desired within the region adjacent to the subsurface motion. This mechanism has been extensively analyzed over only a spatial region on the order of the instability wavelength along the fluid-structure interface. Here we uncover fundamental relations between the behavior of flow instabilities and the frequency response characteristics of the phononic subsurface structure admitting the elastic motion. These relations are then utilized to demonstrate the possibility of extensive spatial expansion of the control regime along the downstream direction with minimal loss of performance—potentially covering the entire surface exposed to the flow.
\end{abstract}

\maketitle
\section{Introduction} \label{sec:Intro}
The interaction between a solid surface and a fluid flow represents a dynamical process that is key to controlling skin-friction drag, flow transition, and flow separation on the surface of air, sea, and land vehicles and numerous other applications including turbomachinery.~\cite{Gad2003,Tiainen2017} The intensity of the skin-friction drag, especially for streamlined bodies, is among the major factors that determine fuel efficiency$-$the lower the drag the higher the fuel efficiency. Skin-friction drag is reduced significantly by the delay of laminar-to-turbulent flow transition.~Flow separation, on its part, is critical to aerodynamic/hydrodynamic stability, form drag, vehicle maneuverability, and applications that involve chemical mixing. Reduction or delay of these quantities are therefore prime objectives in flow control. Key elements that influence these fundamental flow transformations are unstable flow disturbances, or perturbations, such as Tollmien–Schlichting (TS) waves.~\cite{schlichting2016boundary} The linear nature of these waves provides an opportunity to apply phased control by an external stimulus to impede or enhance the growth of these waves by wave superposition.~\cite{milling1981tollmien} Realization of this approach by various active techniques has been pursued in numerous investigations.~\cite{liepmann1982control,liepmann1982active,thomas1983control,Joslin_1995,Grundmann_2008,Amitay_2016} However, these techniques require energy input as well as complex sensing and actuation devices, especially if applied adaptively.~\cite{hu1994feedback,bewley1998optimal} Without closed-loop control, phase locking of a target instability is also required to control the timing of the intervention,~\cite{Amitay_2016} thus limiting the ability to manipulate multiple and spontaneously generated instabilities. \\
\indent To overcome these drawbacks, precise, passive, and responsive/adaptive control of flow instabilities has been demonstrated using a phononic subsurface (PSub).~\cite{Hussein_2015} A PSub comprises a finite phononic structure placed ``underneath" the fluid-structure interface and extending to the interface itself. Given its finite size, it represents a truncated~\cite{davis2011analysis,al2023theory,rosa2022material} phononic material~\cite{hussein2014dynamics,Phani_2017} such as a Bragg scattering phononic crystal~\cite{Hussein_2015,Barnes_2021} or a locally resonant elastic metamaterial.~\cite{kianfar2023phononicNJP} An instability traveling within the flow excites this interface, triggering elastic waves in the PSub which reflect and return back to the flow. The unit-cell and finite-structure characteristics of the PSub may be designed to passively enforce the returning waves to resonate and be out of phase when reentering the flow, causing significant destructive interferences of the continuously incoming flow waves near the surface and subsequently their attenuation over the spatial region covered by the fluid-PSub interface. 
The outcome in this scenario is a local reduction in the skin-friction drag. 
Destabilization, to accelerate the transition to turbulence and possibly delay separation or enhance chemical mixing, is also possible, where, in contrast, the PSub is configured to induce constructive interferences.~\cite{Hussein_2015} In both the stabilization and destabilization cases, the tuning of the PSub requires knowledge of only the frequency, wavelength, and overall modal characteristics of the flow instability.\\
\indent Extensive analysis of PSubs has been done on configurations covering a characteristically narrow spatial domain, spanning a streamwise distance on the order of the wavelength of the target flow instability wave or lower. \cite{Hussein_2015,Barnes_2021,kianfar2023phononicNJP} While installation of only a single PSub demonstrates the fundamental proof of concept, clearly it limits the ultimate overall gains in reducing the skin friction or alternatively accelerating flow transition. As stated in earlier references,~\cite{Hussein_2015,kianfar2023phononicNJP} the local nature of flow-PSub interaction (FPI) naturally facilitates the employment of numerous PSubs to cover an extended region. The demonstrated possible rapid recovery of the kinetic energy of an instability downstream of a PSub ascertains this aspect.~\cite{kianfar2023phononicNJP} Guided by the PSubs design theory,~\cite{Hussein_2015} an array of thin PnC-based PSubs separated with gaps was recently implemented in the context of linearized Navier Stokes simulations.~\cite{michelis2023attenuation}\\
\indent In this article, we uncover a series of fundamental relations between key features in the frequency response of a PSub on the one hand and key features in the responding flow instability field on the other. These relations reveal consequential tradeoffs in the PSub local performance, which must be understood to enable sustained local flow stabilization or destabilization over an extended spatial domain along the downstream direction when a series of repetitively arranged PSubs is applied. Utilizing this information, placement of multiple adjacent PSubs is investigated and the corresponding overall spatial performance is analyzed. The relative size of the PSub interactive surface compared to the instability wavelength and the type of connection of the PSub to the rest of the wall are also studied.
\begin{figure*}[t!]
\centering
\includegraphics{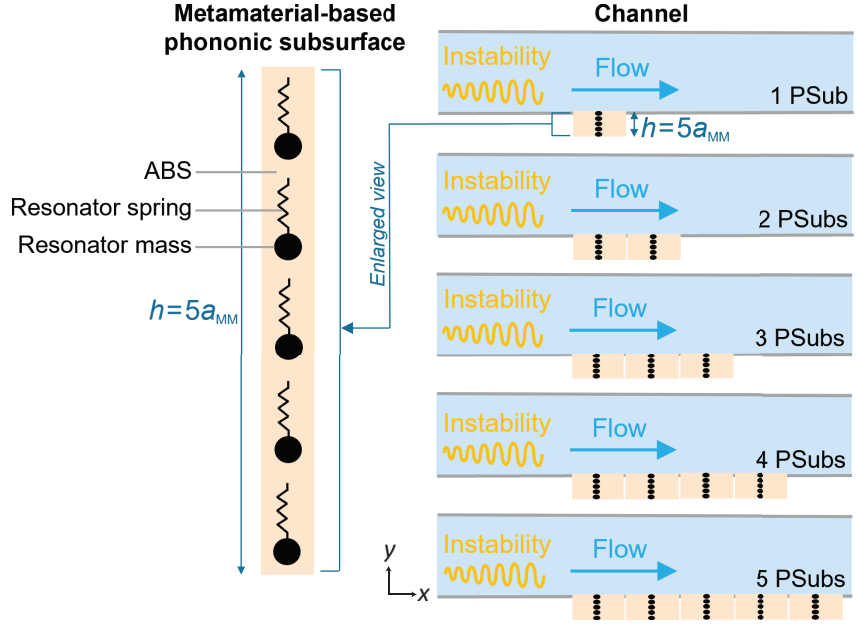}
\caption{\label{fig:fig01} 
Schematic of channel flow with a PSub or more than one PSub installed. Each PSub is composed of a truncated locally resonant elastic metamaterial realized in the form of a rod with a periodic array of spring-mass resonators as shown in the enlarged view schematic on the right.~\cite{kianfar2023phononicNJP} Flow instabilities, e.g. TS waves, will excite the PSub(s) at the top edge(s) (i.e., at the fluid-structure interface), and the PSub(s), in turn, will respond at or near structural resonance and out of phase (for stabilization) with respect to the excitation. This passive response will repeat and manifest in the steady state causing sustained attenuation of reoccurring and continuously incoming instability waves.  The PSub could alternatively be designed to ensure destabilization instead of stabilization by producing an in-phase elastic response.}
\end{figure*}
\begin{figure*}[t!]
\centering
\includegraphics[width=1\textwidth]{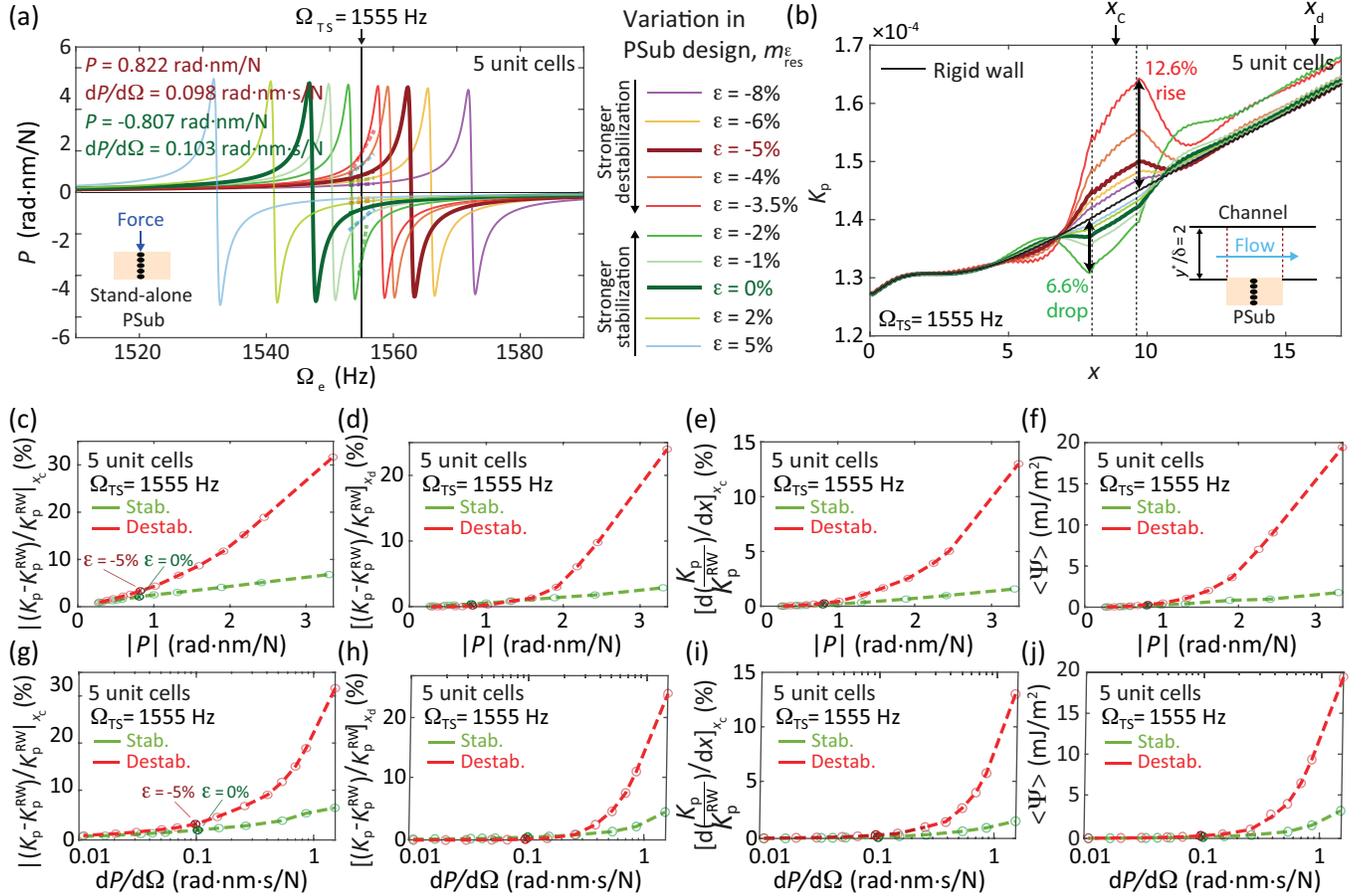}
\caption{\label{fig:fig02} 
Characterization of fundamental relations between PSub performance metric properties and actual flow control performance. Different PSub designs are realized by varying the resonator mass: (a) Performance metric $P$ as a function of the excitation frequency (standalone analysis done prior to the coupled fluid-structure simulations). An inclined line representing the slope $dP/d\Omega$ is shown for each case at the target TS wave frequency. The inset shows a single PSub prior to installation, where each circle represents a mass-spring oscillator. (b) The kinetic energy of flow instability $K_\mathrm{p}$ as a function of the streamwise position for various PSubs (and all-rigid-wall case for comparison), targeting TS wave of frequency $\Omega_{\rm TS}=1555$ Hz. The two vertical dashed lines represent the streamwise leading and trailing edges of the PSub control region at $x_s$ and $x_e$, respectively. The inset illustrates the PSub installation in the channel. Dependence of various flow quantities as a function of $P({\Omega_\mathrm{TS})}$ at the target TS wave frequency: Normalized $K_\mathrm{p}$ difference between PSub and all-rigid-wall case evaluated at (c) $x_c$ and (d) $x_d$. (e) Spatial (streamwise) derivative of $K_\mathrm{p}$  normalized by the rigid wall case evaluated at $x_c$. (f) Total elastodynamic energy of the PSub. Corresponding results are shown for $\mathrm{d}P({\Omega_\mathrm{TS})}/\mathrm{d} \Omega$ dependence in (g)-(j).}
\end{figure*}
\\
\section{Models and Methods} \label{sec: Model}
\indent Governed by the three-dimensional Naiver-Stokes equations, we run a series of direct numerical simulations (DNS) for incompressible channel flows. The velocity vector solution is expressed as ${\bf u}(x,y,z,t)=(u,v,w)$ with components in the streamwise $x$, wall-normal $y$, and the spanwise $z$ directions, respectively, where $t$ denotes time. We run the DNS for a Reynolds number of $Re=\rho_\mathrm{f} U_{\rm c}\delta/\mu_\mathrm{f}=7500$ based on a centerline velocity $U_{\rm c}=17.12$ $\mathrm{m/s}$ and a half-height of the channel $\delta=4.38\times10^{-4}$ $\mathrm{m}$. Liquid water is considered with a density of $\rho_\mathrm{f}=1000$  $\mathrm{kg/m^3}$ and dynamic viscosity of $\mu_\mathrm{f}=1\times10^{-3}$ $\mathrm{kg/ms}$. All subsequent quantities in this study, unless mentioned explicitly, are normalized by the channel's velocity $U_{\rm c}$ and length $\delta$ scales. The channel size is $0 \leq x \leq 30$, $0 \leq y \leq 1$, and $0 \leq z \leq 2 \pi$ for the streamwise, wall-normal, and spanwise directions, respectively. At the inlet of the channel, we superimpose a fully developed Poiseuille flow with an unstable TS mode obtained from linear hydrodynamic stability analysis governed by the Orr-Sommerfeld equation \cite{Orr1907,Sommerfeld1909} and solved for the same $Re$. We select the least-attenuated spatial eigensolution which has a complex wavenumber $\alpha=1.0004-\mathrm{i}0.0062$ and a real non-dimensional frequency $\omega_\mathrm{TS}=0.25$. Following dimensional analysis, the frequency of the TS wave is $\Omega_\mathrm{TS}={\omega_\mathrm{TS}}{U_{\rm c}}/{2\pi}{\delta}= 1555$ $\mathrm{Hz}$. To ensure outgoing waves on the other side of the channel, the disturbances are smoothly brought to zero by attaching a non-reflective buffer region at the outlet. \cite{Dana91,Saiki93,Kucala14} Periodic boundary conditions are applied in the spanwise direction. At the top and bottom walls, rigid no-slip/no-penetration boundary conditions are applied, except within the control region from $x_\mathrm{s}$ to $x_\mathrm{e}$ in the streamwise direction where the rigid wall is replaced by a PSub, or an array of PSubs, at the bottom wall (see Fig.~\ref{fig:fig01}). Each PSub covers the full spanwise width of the channel. Within the control region, the FPI coupling is enforced by means of transpiration boundary conditions~\cite{Lighthill_1958,Sankar_1981,Hussein_2015} (see Sec.~\ref{sec:appendix1}). A variety of flow quantities are calculated to follow; these are defined in Sec.~\ref{sec:appendix2}.\\
\indent The PSub is modeled as a finite linear elastic metamaterial consisting of five rod unit cells with a local mass-spring resonator attached to the middle of each unit cell.~\cite{kianfar2023phononicNJP} The PSub is free to deform at the edge interfacing with the flow (top) and is fixed at the other end (bottom). In our default configuration, we allow every individual PSub to deform in complete independence from the adjacent rigid wall and from the motion of neighboring PSubs, thus its top surface deformation takes a uniform profile across the fluid-PSub interface region.~\cite{Hussein_2015,kianfar2023phononicNJP} The length of the unit cell along the wall-normal direction is $L_\mathrm{UC}= 1$ $\mathrm{cm}$ (i.e., total PSub length is $5$ $\mathrm{cm}$). The resonator frequency is set to $\Omega_\mathrm{res}=2000$ $\mathrm{Hz}$ by tuning the resonator point mass to be ten times heavier than the total mass of the unit-cell base ($m_\mathrm{res}=10\times \rho L_\mathrm{UC}$, where $\rho$ is the base material density). Hence, the stiffness of the resonator spring is $k_\mathrm{res}=m_\mathrm{res}(2\pi \Omega_\mathrm{res})^2$. The base is composed of ABS polymer with a density of $\rho=1200$ $\mathrm{kg/m^3}$ and Young's modulus of $E=3$ $\mathrm{GPa}$. Material damping for the ABS polymer is modeled as viscous proportional damping with constants $q_1=0$ and $q_2=6\times 10^{-8}$.~\cite{kianfar2023phononicNJP} \\
\indent The Navier-Stokes equations are integrated using a time-splitting scheme~\cite{Dana91,Saiki93,Kucala14} on a staggered structured grid system. A two-node iso-parametric finite-element (FE) model is used for determining the PSub nodal axial displacements, velocities, and accelerations~\cite{HusseinJSV06} where time integration is implemented  simultaneously with the flow simulation using an implicit Newmark algorithm \cite{Newmark}. Since the equations for the fluid and the PSub are inverted separately
in the coupled simulations, a conventional serial staggered scheme~\cite{Farhat_2000} is implemented
to couple the two sets of time integration. More details on the computational models and numerical schemes used are detailed in Kianfar and Hussein.~\cite{kianfar2023phononicNJP} \\
 \begin{figure*}[t!]
\centering
\includegraphics[width=1\textwidth]{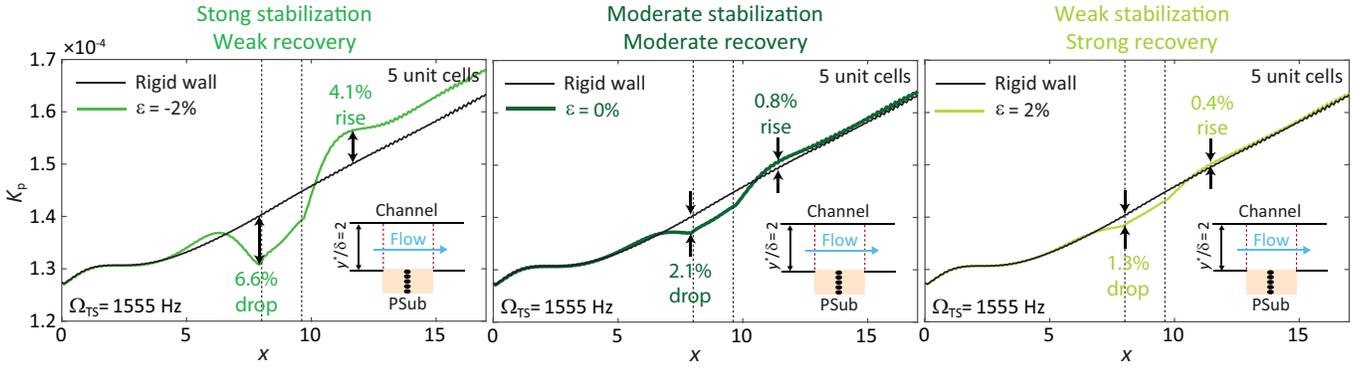}
\caption{\label{fig:fig03} Close-up of three separate cases from among those examined in Fig.~\ref{fig:fig02}, namely strong, moderate, and weak stabilization cases. Strong stabilization (left) enables a significant drop in the kinetic energy of the instability, especially near the PSub leading edge, but comes at the expense of a poor recovery to the reference rigid-wall case level in the downstream region. Weak stabilization (right), on the other hand, offers good donwstream recovery but at the expense of a relatively weak drop in the kinetic energy within the PSub region. Moderate stabilization (middle) provides a compromise, i.e., a moderate drop and a moderate recovery.} 
\end{figure*}
\begin{figure*}[t!]
\centering
\includegraphics{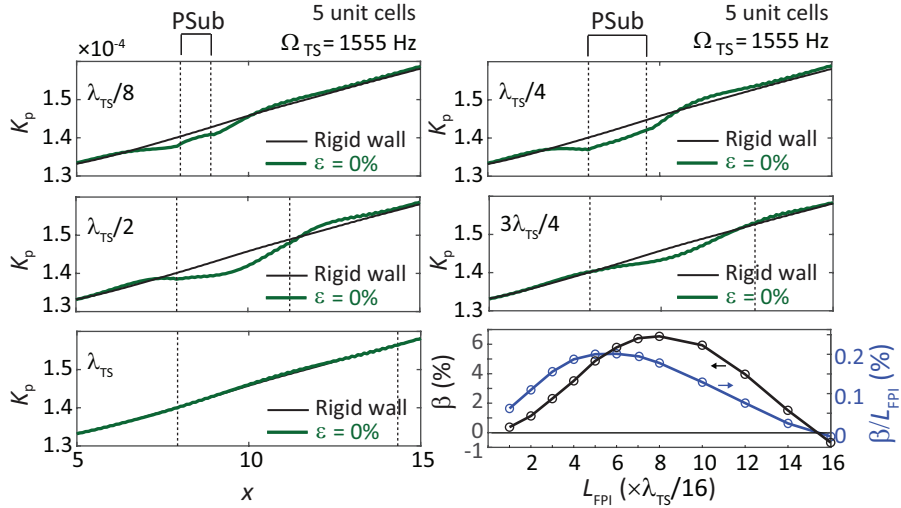}
\caption{\label{fig:fig04} Effect of PSub streamwise length $L_\mathrm{FPI}$ on stabilization intensity, with the length expressed in terms of the TS wave wavelength $\lambda_\mathrm{TS}$. Profile of $K_\mathrm{p}$ as a function of the streamwise position is shown in the first five panels as labeled, and the absolute efficiency metric $\beta$ and relative efficiency metric $\beta/L_\mathrm{FPI}$ of the PSub as a function of $L_\mathrm{FPI}$ are plotted in the bottom right panel.~These results are for an $\varepsilon=0\%$ PSub design with $P({\Omega_{\mathrm{TS}}}) = -0.807$.} 
\end{figure*}
\begin{figure*}[t!]
\centering
\includegraphics[width=0.6\textwidth]{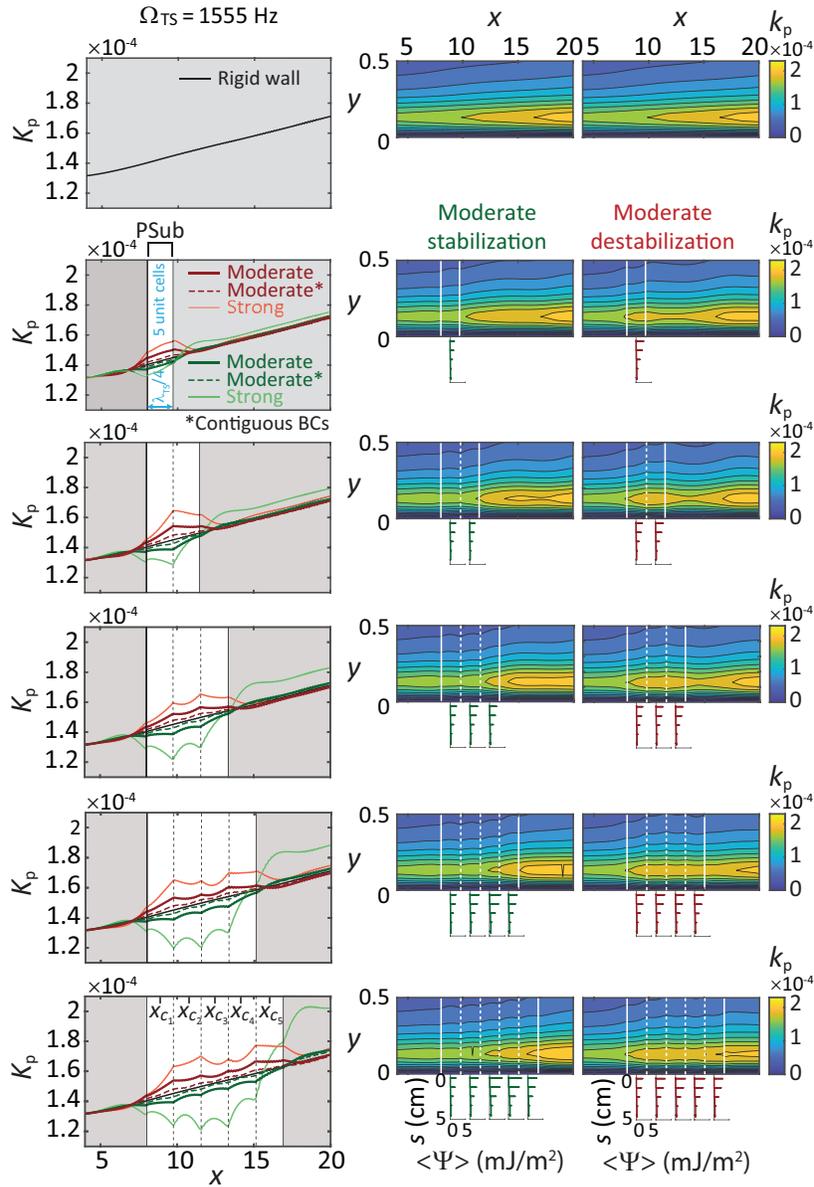}
\caption{\label{fig:fig05} Extension of the passive control surface region using up to five adjacently arranged PSubs. The first column is $K_\mathrm{p}$ as a function of $x$. The perturbation kinetic energy contour in the $x-y$ plane $k_\mathrm{p}$ and the PSub elastodynamic energy $\langle \Psi \rangle$ are shown for stabilization (second column) and destabilization (third column). The horizontal lines indicate the total energy level of the resonating masses.~These results are for the $\varepsilon=0\%$ (moderate stabilization), $\varepsilon=-5\%$ (moderate destabilization),  $\varepsilon=-2\%$ (strong stabilization), and $\varepsilon=-3.5\%$ (strong destabilization) PSub designs, respectively. The moderate control cases under contiguous boundary conditions are shown as well in the first column.}
\end{figure*}
\begin{figure*}[t!]
\centering
\includegraphics{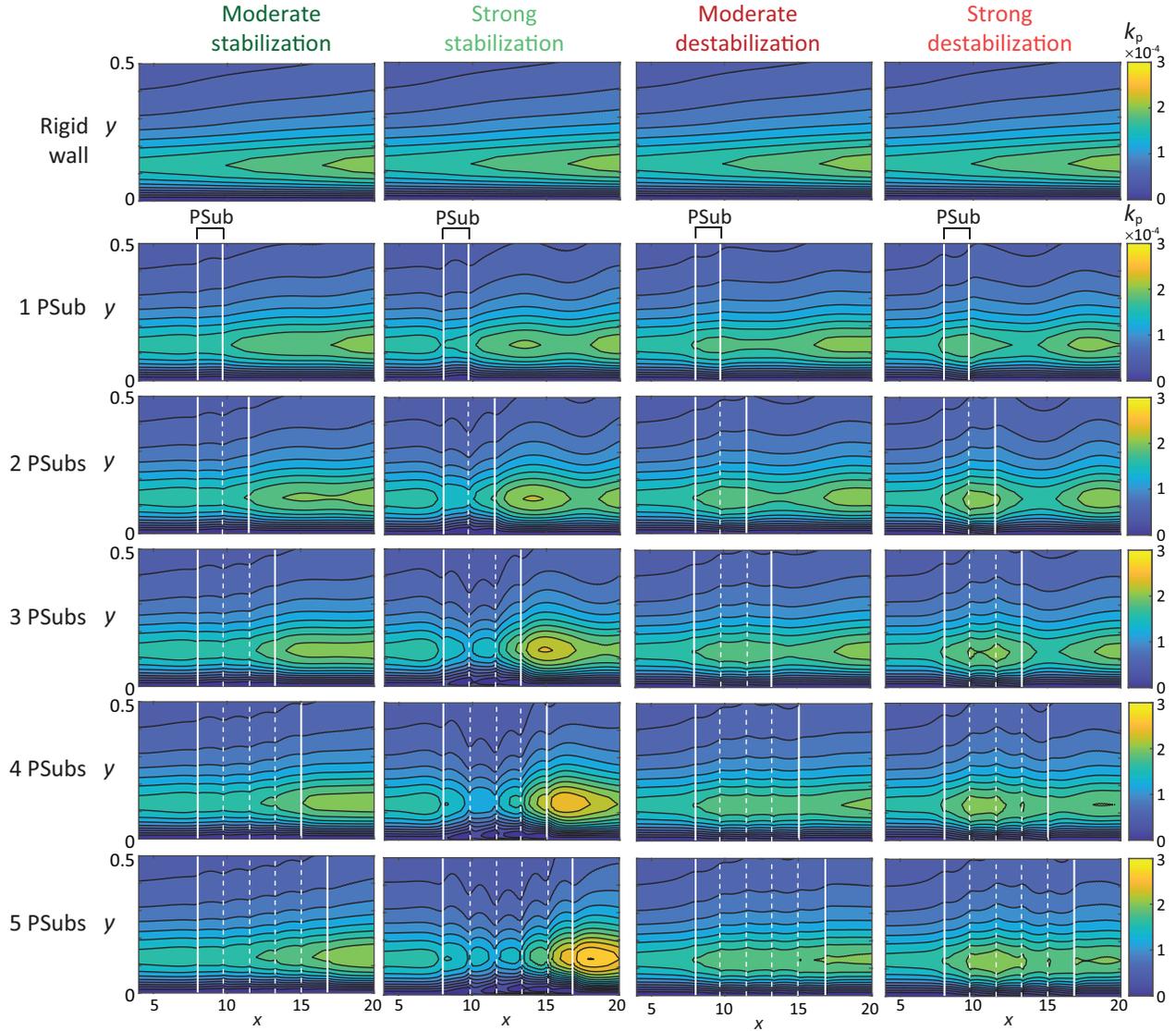}
\caption{\label{fig:fig06} Spatial contours of the perturbation kinetic energy for the selected cases of strong and moderate stabilization or destabilization. The spatial domain of the PSub control regime extends with every added PSub. The moderate control case demonstrates recovery of the downstream kinetic energy to that of the reference rigid-wall case. In contrast, the strong control case displays poor downstream recovery.}
\end{figure*} 
\begin{figure}[t!]
\centering
\includegraphics{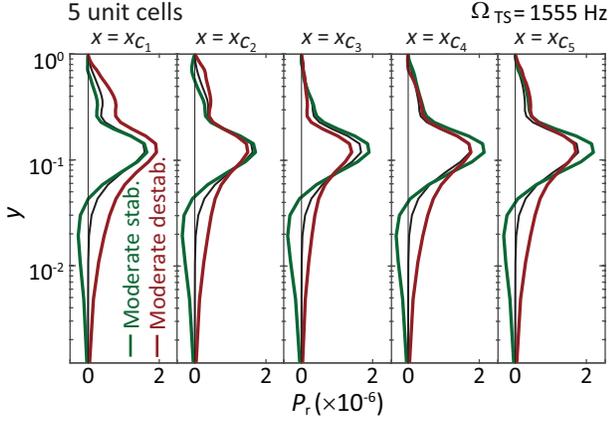}
\caption{\label{fig:fig07} Production rate of the perturbation kinetic energy $P_r$ as a function of the wall-normal coordinate $y$ at the center of each PSub for the extended control configuration comprising five adjacent PSubs.~The $x_{\mathrm{C}_i}$ locations are indicated in the bottom of the first column in Fig.~\ref{fig:fig05}.~These results are for the $\varepsilon=0\%$ (moderate stabilization) and $\varepsilon=-5\%$ (moderate destabilization) PSub designs, respectively.}
\end{figure}
\begin{figure}[t!]
\centering
\includegraphics{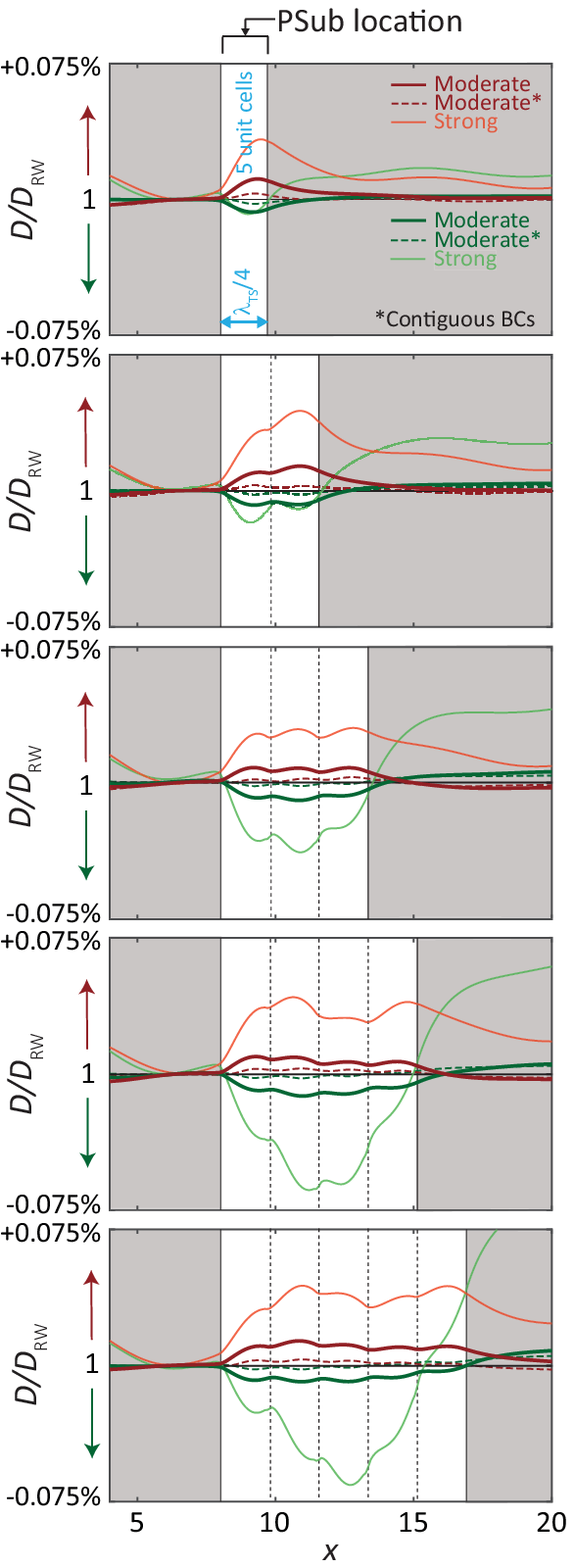}
\caption{\label{fig:fig08} Skin friction drag normalized by the rigid wall case over the extended control region with a number of PSubs installed varying from 1 to 5. The uncontrolled region is shaded in grey.~These results are for the $\varepsilon=0\%$ (moderate stabilization), $\varepsilon=-5\%$ (moderate destabilization),  $\varepsilon=-2\%$ (strong stabilization), and $\varepsilon=-3.5\%$ (strong destabilization) PSub designs, respectively. The moderate control cases under contiguous boundary conditions are shown as well.}
\end{figure}
\section{PSubs performance characterization} 
\label{sec: PSubsP}
The frequency-dependent performance metric $P$ is the primary quantity for assessing the control properties of a PSub. This quantity, introduced by Hussein et al.,~\cite{Hussein_2015} is obtained from the harmonic forced response of the PSub as a standalone structure, prior to coupling with the flow. It is defined as the product of the displacement amplitude (per unit force) and phase of the frequency response function at the PSub surface exposed to the flow due to an excitation at the same location. In Fig.~\ref{fig:fig02}, the response of the flow instability at $\Omega_\mathrm{TS}$ is examined in relation to the PSub $P$ value as well as its frequency derivative $\mathrm{d}P/\mathrm{d} \Omega$. A series of PSub designs are considered with a gradually varying value of the resonator mass such that $m_{\mathrm{res}}^{\epsilon}=m_\mathrm{res}(1+\epsilon)$, reaching up to $\epsilon=\pm 20\%$. This variation shifts the sub-hybridization resonance (SHR) frequency;~\cite{kianfar2023phononicNJP} increasing $m_{\mathrm{res}}^{\epsilon}$ lowers the hybridization band gap which, by extension, lowers the SHR frequency. Moreover, the value and sign of $P({\Omega_\mathrm{TS})}$ and the value of $\mathrm{d}P({\Omega_{\mathrm{TS}}})/\mathrm{d} \Omega$ vary with these changes in  $m_{\mathrm{res}}^{\epsilon}$. As can be seen in Figs.~\ref{fig:fig02}(a) and \ref{fig:fig02}(b) and in earlier studies,~\cite{Hussein_2015,kianfar2023phononicNJP} a positive or negative $P({\Omega_\mathrm{TS})}$ value corresponds to a local stabilization or destabilization effect, respectively, within the control region in the flow. Furthermore, the absolute value $|P|$ correlates with the strength of the stabilization or destabilization effect. The flow response is measured by the change of the wall-normal integral of the perturbation (instability) kinetic energy $K_\mathrm{p}(x)$ compared to the all-rigid-wall case. A rise in $K_\mathrm{p}(x)$ corresponds to destabilization, and vice versa for a drop in $K_\mathrm{p}(x)$. Given that an individual PSub responds to the excitation of a flow instability, it acts as a passively ``responsive" actuator. Thus regardless of its location compared to the phasing of the incoming flow wave at any instant in time, it will respond according to the sign of performance metric$-$causing a stabilizing effect if $P({\Omega_\mathrm{TS})}<0$ and a destabilizing effect if $P({\Omega_\mathrm{TS})}>0$. The responsiveness of a PSub regardless of location is demonstrated further in Sec.~\ref{sec:appendix3}.\\  
\indent In Fig.~\ref{fig:fig02}(b), we observe a maximum change of $K_\mathrm{p}(x)$ of 6.6\% for the strongest stabilization case and 12.6\% for the strongest destabilization case. Higher changes in $K_\mathrm{p}(x)$ may be further realized for PSubs designed to exhibit larger values of $P$ at the instability frequency.~\cite{Hussein_2015} However, it is also observed from Fig.~\ref{fig:fig02}(b) that stronger control within the PSub region appears to take effect at the expense of poorer recovery of the kinetic energy curve downstream of the PSub when compared to the reference rigid-wall case. This behavior is examined more closely in Fig.~\ref{fig:fig03} which highlights the $K_\mathrm{p}(x)$ curve for three of the cases considered in Fig.~\ref{fig:fig02}, representing strong, moderate, and weak stabilization, respectively. In each case, the instability kinetic energy maximum drop at the leading edge of the PSub and maximum rise downstream of the PSub are marked. It is observed that while the drop decreases with weaker stabilization, the   quality of recovery improves, and does so at a favorable rate. This is seen by noting that the ratio of maximum rise to maximum drop in the $K_\mathrm{p}(x)$ curve compared to the rigid-wall behavior is 62.1\%, 38.1\%, and 30.8\% for the strong, moderate, and weak stabilization cases, respectively.   \\
\indent In Fig.~\ref{fig:fig02}(c), the absolute value change in the perturbation kinetic energy due to adding the PSub, $\left(K_\mathrm{p}-K_\mathrm{p}^\mathrm{RW}\right)/K_\mathrm{p}^\mathrm{RW}$, is plotted versus $|P|$ at $\Omega_\mathrm{TS}$ when evaluated at the center of the PSub, $x=x_\mathrm{c}$. The same quantity is plotted in Fig.~\ref{fig:fig02}(d) but at position $x=x_\mathrm{d}$, which is two TS wavelengths downstream of the end of the control region. Throughout this study, the superscript $(.)^\mathrm{RW}$ corresponds to the all-rigid-wall simulations. We observe similar rising trends between Figs.~\ref{fig:fig02}(c) and~\ref{fig:fig02}(d). This more generally quantifies the inherent trade-off noted above for the PSub flow control: the stronger the PSub effect is within the control region, the weaker its recovery downstream of the control region. However, the results show that the influence within the control region can reach about an order of magnitude higher than the influence downstream of the control region, which ascertains the local nature of the control regime. Yet, for relatively large values of $P({\Omega_\mathrm{TS})}$ ($\approx2$ rad$\cdot$nm/N), which occurs when $\Omega_\mathrm{TS}$ closely approaches the SHR frequency, the level of recovery deteriorates. It is also noted that for the same value of $P({\Omega_\mathrm{TS})}$, destabilization always takes place more intensely than stabilization, which is expected because the default state of the unstable flow wave is destabilization; an installed PSub with a positive $P({\Omega_\mathrm{TS})}$ value further enhances this state of destabilization. \\
\indent In Fig.~\ref{fig:fig02}(e), we observe a direct correlation between the spatial growth rate of $K_\mathrm{p}$ within the PSub control region, i.e., $\mathrm{d} K_\mathrm{p}/\mathrm{d}x$, and $P({\Omega_\mathrm{TS})}$. It is noticeable that the growth rate of $K_\mathrm{p}$ is positive within the PSub control region for both the stabilization and destabilization cases. Finally, the level of PSub elastodynamic energy $\Psi(s, t)$ as a function of $P({\Omega_\mathrm{TS})}$ is examined.~As shown in Fig.~\ref{fig:fig02}(f), the time-averaged elastodynamic energy $\langle \Psi \rangle$ in the PSub increases with $P({\Omega_\mathrm{TS})}$ for both the stabilization and destabilization cases. For large $P({\Omega_\mathrm{TS})}$ values, the displacement amplitude of the PSub is correspondingly large, which produces a high amount of strain energy. Here we note that destabilization of the flow perturbation generates more elastodynamic energy compared to stabilization.~\cite{Hussein_2015,kianfar2023phononicNJP} This is because  destabilization corresponds to constructive interference conditions across both the PSub and flowing instability field; while in contrast stabilization corresponds to destructive interference conditions. \\
\indent While $P({\Omega_\mathrm{TS})}$ provides a direct indication of the type and magnitude of the control effect, we observe from Figs.~\ref{fig:fig02}(g)-(j) that the slope $\mathrm{d}P({\Omega_\mathrm{TS})}/\mathrm{d}\Omega$ also provides clear correlations with the response of the flow instability field. Compared to the trends with respect to $P({\Omega_\mathrm{TS})}$, the Figs.~\ref{fig:fig02}(g)-(j) trends show more abrupt changes at high values of $\mathrm{d}P({\Omega_\mathrm{TS})}/\mathrm{d}\Omega$. Further analysis summarized in Sec.~\ref{sec:appendix4} reveals that the $\mathrm{d}P({\Omega_\mathrm{TS})}/\mathrm{d}\Omega$ metric provides additional details on the behavior of $K_\mathrm{p}$, e.g., two PSubs with the same value of $P({\Omega_\mathrm{TS})}$ will have similar performance, but with subtle differences that correlate precisely with $\mathrm{d}P({\Omega_\mathrm{TS})}/\mathrm{d}\Omega$.  \\
\indent Next we study the effect of the relative length of the PSub interface with the flow.  We examine installing a single PSub with an FPI length larger than the TS wavelength. We observe in Fig.~\ref{fig:fig04}(a) for the $\epsilon=0\%$ design that the curvature and overall profile of $K_\mathrm{p}$ change when the normalized FPI length $L_\mathrm{FPI}$ is varied with respect to the TS wavelength. Within the boundaries of the PSub control region, the trend of $K_\mathrm{p}$ with respect to $x$ is concave for $L_\mathrm{FPI} \leq \lambda_\mathrm{TS}/4$, roughly linear for $L_\mathrm{FPI} = \lambda_\mathrm{TS}/4$, and convex for $L_\mathrm{FPI} > \lambda_\mathrm{TS}/4$. These results are concisely quantified by an efficiency metric defined as $\beta = \int_{x_s}^{x_e} {\left(K_\mathrm{p} - K^\mathrm{RW}_\mathrm{p} \right)}/{K^\mathrm{RW}_\mathrm{p} } dx$, which is plotted in the bottom-right panel of Fig.~\ref{fig:fig04}. This metric provides an overall performance measure across the entire control region of the PSub, spanning the distance $x_s \leq x \leq x_e$ and normalized pointwise by the corresponding perturbation kinetic energy of the rigid-wall case. We observe that the reduction of $K_\mathrm{p}$ asymptotically vanishes when $L_\mathrm{FPI}$ approaches either zero (rigid wall) or one full TS wavelength. This indicates that an increase in the downstream length of a single PSub is not a viable option for increasing the spatial extension of the control region, thus placement of an array of adjacent PSubs is needed. The results of Fig.~\ref{fig:fig04} indicate that the optimal $L_\mathrm{FPI}$ value for the strength of intervention lies between $(1/4)$th and $(3/8)$th of a TS wavelength when considering the relative efficiency metric $\beta/L_\mathrm{FPI}$. In the following analysis, we consider arrays of adjacent PSubs with $L_\mathrm{FPI} = \lambda_\mathrm{FPI}/4$ selected for each PSub.
\section{Performance of arrays of adjacent PSubs} \label{sec: PSubsAd}
\indent In this section, we investigate the effect of installing multiple PSubs to extend the spatial range of flow control. We consider primarily the two ``moderate" cases, namely the $\epsilon=0\%$ PSub design (for stabilization) and the $\epsilon=-5\%$ PSub design (for destabilization).~These two cases are highlighted in Figs.~\ref{fig:fig02}(a) and~\ref{fig:fig02}(b) with thicker curves. These two designs exhibit relatively close absolute values of the performance metric and its slope:  [$P({\Omega_\mathrm{TS})} = -0.807$ rad$\cdot$nm/N; $\mathrm{d}P({\Omega_{\mathrm{TS}}})/\mathrm{d}\Omega = 0.103$ rad$\cdot$nm$\cdot$s/N] and [$P({\Omega_\mathrm{TS})} = 0.822$ rad$\cdot$nm/N; $\mathrm{d}P({\Omega_{\mathrm{TS}}})/\mathrm{d}\Omega = 0.098$ rad$\cdot$nm$\cdot$s/N], respectively. This selection offers a favorable trade-off between the strength of stabilization/destabilization and the level of recovery downstream of the PSub, which enables the use of multiple adjacent PSubs to allow for a substantial spatial extension of the control regime.\\
\indent Figure \ref{fig:fig05} presents the time-averaged $x$-dependent profile of $K_\mathrm{p}(x)$ (first column) and $x$- and $y$-dependent contour of the perturbation kinetic energy $k_\mathrm{p}(x,y)$ (second and third columns) for these selected stabilization and destabilization cases.~The corresponding time-averaged elastodynamic energy $\langle \Psi \rangle$ along the domain of each PSub is also plotted in conjunction with the flow instability contours. As shown, five different PSub assemblies are considered, comprising 1, 2, 3, 4, or 5 PSubs arranged adjacent to each other.~Each PSub is free to move independently of its adjacent PSub(s). For the rigid wall case (the first row in Figure \ref{fig:fig05}), there is a peak region of $k_\mathrm{p}$  intensity at nearly $y=0.2$ close to the wall. As more PSubs are installed, we observe the influence extending further downstream, achieving the desired control for both the stabilization and destabilization cases.  For the stabilization case, as more PSubs are introduced the peak $k_\mathrm{p}$ region is rendered smaller as it is pushed further downstream (i.e., smaller yellow region, where yellow represents higher perturbation intensity). The opposite effect is observed for the destabilization cases. In summary, the intensities and slopes of the performance metric within the control region and the corresponding intensities downstream of it are all consistent with the performance metric relations shown in Fig.~\ref{fig:fig02} which are obtained without conducting flow simulations. The intensity of $\langle \Psi \rangle$ for the single- and two-PSub assemblies is seen to be higher for the destabilization case compared to the stabilization case, consistent with in-phase and out-of-phase behaviors as mentioned earlier.~\cite{kianfar2023phononicNJP} The $\langle \Psi \rangle$ behaviors are more complex when more than two PSubs are installed. A key observation from these results is that the desired control function (stabilization or destabilization) is realized within the control region (i.e., the white space in the left panel of Fig.~\ref{fig:fig05}) with generally good recovery downstream of the last PSub in the array. We also observe that when several PSubs are employed the performance peaks somewhere in the interior of the spatial domain covered by the PSubs. \\
\indent To appreciate the implications of the strength-recovery tradeoff, in Fig.~\ref{fig:fig05} we superimpose an additional pair of cases corresponding to ``strong" performance; these are $\epsilon=-2\%$ for stabilization and $\epsilon=-3.5\%$ for destabilization. For these extreme cases, we see significantly stronger stabilization or destabilization within the control region at the expense of a poor downstream recovery (as expected), but, moreover, we observe that not all the region covered by the PSubs (the white region in the figure) displays the desired control. For example, for the 5-PSubs strong stabilization case, the instability kinetic energy exceeds that of the rigid-wall case in the region corresponding to the fifth PSub. The contrast in performance, in terms of both the strength of control within the PSubs region and quality of recovery downstream of the PSubs, is further demonstrated in Fig.~\ref{fig:fig06} where the $x-y$ contour of the instability kinetic energy is plotted for both the moderate and strong pairs of PSub designs. One key consideration, however, is the manner by which a PSub is connected to the rigid wall. If instead of completely free ``elevator-type" type motion is admitted, the PSub is fixed to the rigid wall at each end, the quality of recovery is seen to be significantly improved as demonstrated in Sec.~\ref{sec:appendix5} and also shown in Fig. ~\ref{fig:fig02} for the multiple PSub cases. This, in turn, offers better extensibility of PSubs spatial coverage. \\
 \indent The production rate of perturbation kinetic energy is another key quantity in our investigation as it characterizes the energy transfer into or out of instability waves. \cite{Prandtl1921,Morris_1976,Hussein_2015,kianfar2023phononicNJP} In Fig.~\ref{fig:fig07} the dimensionless production rate $P_{\rm r}(x,y)$ at the center of each PSub is plotted for the case of five adjacent PSubs. For stabilization, we observe $P_{\rm r}$ to drop to negative values compared to the rigid wall case within $x_s \leq x \leq x_c$, which indicates less transfer of energy from the mean flow to the instability.\cite{Hussein_2015,kianfar2023phononicNJP} This behavior takes place in the region very close to the wall, $y < 0.1$, where the influence of the PSub's motion on the flow field is substantial. We notice the strongest negative $P_{\rm r}$ occurs at the center of the third PSub. Selection of a weaker PSub design with better recovery increases the maximum number of PSubs that may be employed for favorable control within the PSubs domain, but this comes at the expense of the  $k_\mathrm{p}$ reduction per PSub.  Downstream of $x_c$, the near-wall negative production rate diminishes; also, away from the wall along the wall-normal direction the peak $P_{\rm r}$ becomes greater than the rigid wall case, showing a faster growth rate for the  instability. Similar but opposite trends are observed for the destabilization case.\\
\indent In Fig.~\ref{fig:fig08}, we demonstrate the impact of PSubs on the normalized friction drag force $D(x)/D_{\rm RW}(x)$, showing a sustained reduction or increase over the extended control region following the trade-off constraints described above. The changes incurred in $D(x)/D_{\rm RW}(x)$ due to the presence of PSubs are small due to the small magnitude of the perturbation wave relative to the mean flow,~\cite{kianfar2023phononicNJP} but still reveal valuable qualitative information.~A practical impact on friction drag will arise when the PSubs collectively cause a delay in transition. By extending the spatial domain of local flow control, the location of the transition point will consequently be delayed. Finally, it should be noted that the time-averaged wall shear stress $\langle \tau_w (x) \rangle$ is dominated by the shear associated with the mean flow. However, with the presence of the PSubs there is a small but noticeable $\langle \hat{u} \hat{v} \rangle$ component at the wall due to the infinitesimal elastic motion of the PSub surface interfacing with the flow.\\
\section{Conclusion}  \label{sec:Concl}
\indent In conclusion, we have investigated using DNS the feasibility of employing multiple phononic-subsurface units for passive control of wall-bounded unstable flows. We demonstrated local flow stabilization and destabilization over an extended spatial domain along the downstream direction. The behavior of a single PSub may be fully characterized offline (i.e., independent of the coupled fluid-structure simulations) by examining \textit{a priori} the properties of the PSub performance metric $P({\Omega_\mathrm{TS})}$ and its frequency derivative $dP({\Omega_\mathrm{TS})}/d \Omega$. We have shown that a single PSub offers the strongest control when its interface length in the downstream direction is between (1/4)th and (3/8)th of the target TS wavelength. Ultimately, the trade-off between the strength of instability control within the PSub control region and the strength of recovery downstream of that region has to be considered to maximize the spatial extent of downstream passive control by the placement of multiple adjacent PSubs. It was also shown the fixing the edges of each PSub to the rigid wall to allow only the interior part to elastically deform leads to weaker control but much improved recovery. Future research may consider mixing-and-matching different types of PSubs to further optimize the underlying strength-recovery trade-off for further extension of spatial coverage.   \\

\section{Acknowledgement}
This work utilized the RMACC Summit supercomputer, which is supported by the National Science Foundation (awards ACI-1532235 and ACI-1532236), the University of Colorado Boulder, and Colorado State University. The Summit supercomputer is a
joint effort of the University of Colorado Boulder and Colorado State University.

\bibliography{RefPSubs}%

\providecommand{\noopsort}[1]{}\providecommand{\singleletter}[1]{#1}%
\begin{thebibliography}{35}%
\makeatletter
\providecommand \@ifxundefined [1]{%
 \@ifx{#1\undefined}
}%
\providecommand \@ifnum [1]{%
 \ifnum #1\expandafter \@firstoftwo
 \else \expandafter \@secondoftwo
 \fi
}%
\providecommand \@ifx [1]{%
 \ifx #1\expandafter \@firstoftwo
 \else \expandafter \@secondoftwo
 \fi
}%
\providecommand \natexlab [1]{#1}%
\providecommand \enquote  [1]{``#1''}%
\providecommand \bibnamefont  [1]{#1}%
\providecommand \bibfnamefont [1]{#1}%
\providecommand \citenamefont [1]{#1}%
\providecommand \href@noop [0]{\@secondoftwo}%
\providecommand \href [0]{\begingroup \@sanitize@url \@href}%
\providecommand \@href[1]{\@@startlink{#1}\@@href}%
\providecommand \@@href[1]{\endgroup#1\@@endlink}%
\providecommand \@sanitize@url [0]{\catcode `\\12\catcode `\$12\catcode
  `\&12\catcode `\#12\catcode `\^12\catcode `\_12\catcode `\%12\relax}%
\providecommand \@@startlink[1]{}%
\providecommand \@@endlink[0]{}%
\providecommand \url  [0]{\begingroup\@sanitize@url \@url }%
\providecommand \@url [1]{\endgroup\@href {#1}{\urlprefix }}%
\providecommand \urlprefix  [0]{URL }%
\providecommand \Eprint [0]{\href }%
\providecommand \doibase [0]{http://dx.doi.org/}%
\providecommand \selectlanguage [0]{\@gobble}%
\providecommand \bibinfo  [0]{\@secondoftwo}%
\providecommand \bibfield  [0]{\@secondoftwo}%
\providecommand \translation [1]{[#1]}%
\providecommand \BibitemOpen [0]{}%
\providecommand \bibitemStop [0]{}%
\providecommand \bibitemNoStop [0]{.\EOS\space}%
\providecommand \EOS [0]{\spacefactor3000\relax}%
\providecommand \BibitemShut  [1]{\csname bibitem#1\endcsname}%
\let\auto@bib@innerbib\@empty
\bibitem [{\citenamefont {Gad-el Hak}, \citenamefont {Pollard},\ and\
  \citenamefont {Bonnet}(2003)}]{Gad2003}%
  \BibitemOpen
  \bibfield  {author} {\bibinfo {author} {\bibfnamefont {M.}~\bibnamefont
  {Gad-el Hak}}, \bibinfo {author} {\bibfnamefont {A.}~\bibnamefont {Pollard}},
  \ and\ \bibinfo {author} {\bibfnamefont {J.-P.}\ \bibnamefont {Bonnet}},\
  }\href@noop {} {\emph {\bibinfo {title} {Flow control: Fundamentals and
  practices}}},\ Vol.~\bibinfo {volume} {53}\ (\bibinfo  {publisher} {Springer
  Science \& Business Media},\ \bibinfo {year} {2003})\BibitemShut {NoStop}%
\bibitem [{\citenamefont {Tiainen}\ \emph {et~al.}(2017)\citenamefont
  {Tiainen}, \citenamefont {Gr{\"o}nman}, \citenamefont {Jaatinen-V{\"a}rri},\
  and\ \citenamefont {Backman}}]{Tiainen2017}%
  \BibitemOpen
  \bibfield  {author} {\bibinfo {author} {\bibfnamefont {J.}~\bibnamefont
  {Tiainen}}, \bibinfo {author} {\bibfnamefont {A.}~\bibnamefont
  {Gr{\"o}nman}}, \bibinfo {author} {\bibfnamefont {A.}~\bibnamefont
  {Jaatinen-V{\"a}rri}}, \ and\ \bibinfo {author} {\bibfnamefont
  {J.}~\bibnamefont {Backman}},\ }\bibfield  {title} {\enquote {\bibinfo
  {title} {Flow control methods and their applicability in
  low-{R}eynolds-number centrifugal compressors—{A} review},}\ }\href@noop {}
  {\bibfield  {journal} {\bibinfo  {journal} {International Journal of
  Turbomachinery, Propulsion and Power}\ }\textbf {\bibinfo {volume} {3}},\
  \bibinfo {pages} {2} (\bibinfo {year} {2017})}\BibitemShut {NoStop}%
\bibitem [{\citenamefont {Schlichting}\ and\ \citenamefont
  {Gersten}(2016)}]{schlichting2016boundary}%
  \BibitemOpen
  \bibfield  {author} {\bibinfo {author} {\bibfnamefont {H.}~\bibnamefont
  {Schlichting}}\ and\ \bibinfo {author} {\bibfnamefont {K.}~\bibnamefont
  {Gersten}},\ }\href@noop {} {\emph {\bibinfo {title} {Boundary-layer
  theory}}}\ (\bibinfo  {publisher} {springer},\ \bibinfo {year}
  {2016})\BibitemShut {NoStop}%
\bibitem [{\citenamefont {Milling}(1981)}]{milling1981tollmien}%
  \BibitemOpen
  \bibfield  {author} {\bibinfo {author} {\bibfnamefont {R.~W.}\ \bibnamefont
  {Milling}},\ }\bibfield  {title} {\enquote {\bibinfo {title}
  {Tollmien--schlichting wave cancellation},}\ }\href@noop {} {\bibfield
  {journal} {\bibinfo  {journal} {The Physics of Fluids}\ }\textbf {\bibinfo
  {volume} {24}},\ \bibinfo {pages} {979--981} (\bibinfo {year}
  {1981})}\BibitemShut {NoStop}%
\bibitem [{\citenamefont {Liepmann}, \citenamefont {Brown},\ and\ \citenamefont
  {Nosenchuck}(1982)}]{liepmann1982control}%
  \BibitemOpen
  \bibfield  {author} {\bibinfo {author} {\bibfnamefont {H.}~\bibnamefont
  {Liepmann}}, \bibinfo {author} {\bibfnamefont {G.}~\bibnamefont {Brown}}, \
  and\ \bibinfo {author} {\bibfnamefont {D.}~\bibnamefont {Nosenchuck}},\
  }\bibfield  {title} {\enquote {\bibinfo {title} {Control of
  laminar-instability waves using a new technique},}\ }\href@noop {} {\bibfield
   {journal} {\bibinfo  {journal} {Journal of Fluid Mechanics}\ }\textbf
  {\bibinfo {volume} {118}},\ \bibinfo {pages} {187--200} (\bibinfo {year}
  {1982})}\BibitemShut {NoStop}%
\bibitem [{\citenamefont {Liepmann}\ and\ \citenamefont
  {Nosenchuck}(1982)}]{liepmann1982active}%
  \BibitemOpen
  \bibfield  {author} {\bibinfo {author} {\bibfnamefont {H.}~\bibnamefont
  {Liepmann}}\ and\ \bibinfo {author} {\bibfnamefont {D.}~\bibnamefont
  {Nosenchuck}},\ }\bibfield  {title} {\enquote {\bibinfo {title} {Active
  control of laminar-turbulent transition},}\ }\href@noop {} {\bibfield
  {journal} {\bibinfo  {journal} {Journal of Fluid Mechanics}\ }\textbf
  {\bibinfo {volume} {118}},\ \bibinfo {pages} {201--204} (\bibinfo {year}
  {1982})}\BibitemShut {NoStop}%
\bibitem [{\citenamefont {Thomas}(1983)}]{thomas1983control}%
  \BibitemOpen
  \bibfield  {author} {\bibinfo {author} {\bibfnamefont {A.~S.}\ \bibnamefont
  {Thomas}},\ }\bibfield  {title} {\enquote {\bibinfo {title} {The control of
  boundary-layer transition using a wave-superposition principle},}\
  }\href@noop {} {\bibfield  {journal} {\bibinfo  {journal} {Journal of Fluid
  Mechanics}\ }\textbf {\bibinfo {volume} {137}},\ \bibinfo {pages} {233--250}
  (\bibinfo {year} {1983})}\BibitemShut {NoStop}%
\bibitem [{\citenamefont {Joslin}\ \emph {et~al.}(1995)\citenamefont {Joslin},
  \citenamefont {Nicolaides}, \citenamefont {Erlebacher}, \citenamefont
  {Hussaini},\ and\ \citenamefont {Gunzburger}}]{Joslin_1995}%
  \BibitemOpen
  \bibfield  {author} {\bibinfo {author} {\bibfnamefont {R.~D.}\ \bibnamefont
  {Joslin}}, \bibinfo {author} {\bibfnamefont {R.~A.}\ \bibnamefont
  {Nicolaides}}, \bibinfo {author} {\bibfnamefont {G.}~\bibnamefont
  {Erlebacher}}, \bibinfo {author} {\bibfnamefont {M.~Y.}\ \bibnamefont
  {Hussaini}}, \ and\ \bibinfo {author} {\bibfnamefont {M.~D.}\ \bibnamefont
  {Gunzburger}},\ }\bibfield  {title} {\enquote {\bibinfo {title} {Active
  control of boundary-layer instabilities: Use of sensors and spectral
  controller},}\ }\href@noop {} {\bibfield  {journal} {\bibinfo  {journal}
  {AIAA Journal}\ }\textbf {\bibinfo {volume} {33}},\ \bibinfo {pages}
  {1521--1523} (\bibinfo {year} {1995})}\BibitemShut {NoStop}%
\bibitem [{\citenamefont {Grundmann}\ and\ \citenamefont
  {Tropea}(2008)}]{Grundmann_2008}%
  \BibitemOpen
  \bibfield  {author} {\bibinfo {author} {\bibfnamefont {S.}~\bibnamefont
  {Grundmann}}\ and\ \bibinfo {author} {\bibfnamefont {C.}~\bibnamefont
  {Tropea}},\ }\bibfield  {title} {\enquote {\bibinfo {title} {Active
  cancellation of artificially introduced tollmien–schlichting waves using
  plasma actuators},}\ }\href@noop {} {\bibfield  {journal} {\bibinfo
  {journal} {Experiments in Fluids}\ }\textbf {\bibinfo {volume} {44}},\
  \bibinfo {pages} {795--806} (\bibinfo {year} {2008})}\BibitemShut {NoStop}%
\bibitem [{\citenamefont {Amitay}, \citenamefont {Tuna},\ and\ \citenamefont
  {Dell’Orso}(2016)}]{Amitay_2016}%
  \BibitemOpen
  \bibfield  {author} {\bibinfo {author} {\bibfnamefont {M.}~\bibnamefont
  {Amitay}}, \bibinfo {author} {\bibfnamefont {B.~A.}\ \bibnamefont {Tuna}}, \
  and\ \bibinfo {author} {\bibfnamefont {H.}~\bibnamefont {Dell’Orso}},\
  }\bibfield  {title} {\enquote {\bibinfo {title} {Identification and
  mitigation of {T-S} waves using localized dynamic surface modification},}\
  }\href@noop {} {\bibfield  {journal} {\bibinfo  {journal} {Physics of
  Fluids}\ }\textbf {\bibinfo {volume} {28}},\ \bibinfo {pages} {064103}
  (\bibinfo {year} {2016})}\BibitemShut {NoStop}%
\bibitem [{\citenamefont {Hu}\ and\ \citenamefont
  {Bau}(1994)}]{hu1994feedback}%
  \BibitemOpen
  \bibfield  {author} {\bibinfo {author} {\bibfnamefont {H.~H.}\ \bibnamefont
  {Hu}}\ and\ \bibinfo {author} {\bibfnamefont {H.~H.}\ \bibnamefont {Bau}},\
  }\bibfield  {title} {\enquote {\bibinfo {title} {Feedback control to delay or
  advance linear loss of stability in planar poiseuille flow},}\ }\href@noop {}
  {\bibfield  {journal} {\bibinfo  {journal} {Proceedings of the Royal Society
  of London. Series A: Mathematical and Physical Sciences}\ }\textbf {\bibinfo
  {volume} {447}},\ \bibinfo {pages} {299--312} (\bibinfo {year}
  {1994})}\BibitemShut {NoStop}%
\bibitem [{\citenamefont {Bewley}\ and\ \citenamefont
  {Liu}(1998)}]{bewley1998optimal}%
  \BibitemOpen
  \bibfield  {author} {\bibinfo {author} {\bibfnamefont {T.~R.}\ \bibnamefont
  {Bewley}}\ and\ \bibinfo {author} {\bibfnamefont {S.}~\bibnamefont {Liu}},\
  }\bibfield  {title} {\enquote {\bibinfo {title} {Optimal and robust control
  and estimation of linear paths to transition},}\ }\href@noop {} {\bibfield
  {journal} {\bibinfo  {journal} {Journal of Fluid Mechanics}\ }\textbf
  {\bibinfo {volume} {365}},\ \bibinfo {pages} {305--349} (\bibinfo {year}
  {1998})}\BibitemShut {NoStop}%
\bibitem [{\citenamefont {Hussein}\ \emph {et~al.}(2015)\citenamefont
  {Hussein}, \citenamefont {Biringen}, \citenamefont {Bilal},\ and\
  \citenamefont {Kucala}}]{Hussein_2015}%
  \BibitemOpen
  \bibfield  {author} {\bibinfo {author} {\bibfnamefont {M.~I.}\ \bibnamefont
  {Hussein}}, \bibinfo {author} {\bibfnamefont {S.}~\bibnamefont {Biringen}},
  \bibinfo {author} {\bibfnamefont {O.~R.}\ \bibnamefont {Bilal}}, \ and\
  \bibinfo {author} {\bibfnamefont {A.}~\bibnamefont {Kucala}},\ }\bibfield
  {title} {\enquote {\bibinfo {title} {Flow stabilization by subsurface
  phonons},}\ }\href {\doibase 10.1098/rspa.2014.0928} {\bibfield  {journal}
  {\bibinfo  {journal} {Proceedings of the Royal Society A}\ }\textbf {\bibinfo
  {volume} {471}},\ \bibinfo {pages} {20140928} (\bibinfo {year}
  {2015})}\BibitemShut {NoStop}%
\bibitem [{\citenamefont {Davis}\ \emph {et~al.}(2011)\citenamefont {Davis},
  \citenamefont {Tomchek}, \citenamefont {Flores}, \citenamefont {Liu},\ and\
  \citenamefont {Hussein}}]{davis2011analysis}%
  \BibitemOpen
  \bibfield  {author} {\bibinfo {author} {\bibfnamefont {B.~L.}\ \bibnamefont
  {Davis}}, \bibinfo {author} {\bibfnamefont {A.~S.}\ \bibnamefont {Tomchek}},
  \bibinfo {author} {\bibfnamefont {E.~A.}\ \bibnamefont {Flores}}, \bibinfo
  {author} {\bibfnamefont {L.}~\bibnamefont {Liu}}, \ and\ \bibinfo {author}
  {\bibfnamefont {M.~I.}\ \bibnamefont {Hussein}},\ }\bibfield  {title}
  {\enquote {\bibinfo {title} {Analysis of periodicity termination in phononic
  crystals},}\ }in\ \href@noop {} {\emph {\bibinfo {booktitle} {ASME
  International Mechanical Engineering Congress and Exposition}}},\
  Vol.~\bibinfo {volume} {8}\ (\bibinfo {year} {2011})\ pp.\ \bibinfo {pages}
  {973--977}\BibitemShut {NoStop}%
\bibitem [{\citenamefont {Al~Ba'ba'a}\ \emph {et~al.}(2023)\citenamefont
  {Al~Ba'ba'a}, \citenamefont {Willey}, \citenamefont {Chen}, \citenamefont
  {Juhl},\ and\ \citenamefont {Nouh}}]{al2023theory}%
  \BibitemOpen
  \bibfield  {author} {\bibinfo {author} {\bibfnamefont {H.~B.}\ \bibnamefont
  {Al~Ba'ba'a}}, \bibinfo {author} {\bibfnamefont {C.~L.}\ \bibnamefont
  {Willey}}, \bibinfo {author} {\bibfnamefont {V.~W.}\ \bibnamefont {Chen}},
  \bibinfo {author} {\bibfnamefont {A.~T.}\ \bibnamefont {Juhl}}, \ and\
  \bibinfo {author} {\bibfnamefont {M.}~\bibnamefont {Nouh}},\ }\bibfield
  {title} {\enquote {\bibinfo {title} {Theory of truncation resonances in
  continuum rod-based phononic crystals with generally asymmetric unit
  cells},}\ }\href@noop {} {\bibfield  {journal} {\bibinfo  {journal} {Advanced
  Theory and Simulations}\ }\textbf {\bibinfo {volume} {6}},\ \bibinfo {pages}
  {2200700} (\bibinfo {year} {2023})}\BibitemShut {NoStop}%
\bibitem [{\citenamefont {Rosa}\ \emph {et~al.}(2022)\citenamefont {Rosa},
  \citenamefont {Davis}, \citenamefont {Liu}, \citenamefont {Ruzzene},\ and\
  \citenamefont {Hussein}}]{rosa2022material}%
  \BibitemOpen
  \bibfield  {author} {\bibinfo {author} {\bibfnamefont {M.~I.}\ \bibnamefont
  {Rosa}}, \bibinfo {author} {\bibfnamefont {B.~L.}\ \bibnamefont {Davis}},
  \bibinfo {author} {\bibfnamefont {L.}~\bibnamefont {Liu}}, \bibinfo {author}
  {\bibfnamefont {M.}~\bibnamefont {Ruzzene}}, \ and\ \bibinfo {author}
  {\bibfnamefont {M.~I.}\ \bibnamefont {Hussein}},\ }\bibfield  {title}
  {\enquote {\bibinfo {title} {Material vs. structure: Topological origins of
  band-gap truncation resonances in periodic structures},}\ }\href@noop {}
  {\bibfield  {journal} {\bibinfo  {journal} {arXiv preprint arXiv:2301.00101}\
  } (\bibinfo {year} {2022})}\BibitemShut {NoStop}%
\bibitem [{\citenamefont {Hussein}, \citenamefont {Leamy},\ and\ \citenamefont
  {Ruzzene}(2014)}]{hussein2014dynamics}%
  \BibitemOpen
  \bibfield  {author} {\bibinfo {author} {\bibfnamefont {M.~I.}\ \bibnamefont
  {Hussein}}, \bibinfo {author} {\bibfnamefont {M.~J.}\ \bibnamefont {Leamy}},
  \ and\ \bibinfo {author} {\bibfnamefont {M.}~\bibnamefont {Ruzzene}},\
  }\bibfield  {title} {\enquote {\bibinfo {title} {Dynamics of phononic
  materials and structures: Historical origins, recent progress, and future
  outlook},}\ }\href@noop {} {\bibfield  {journal} {\bibinfo  {journal}
  {Applied Mechanics Reviews}\ }\textbf {\bibinfo {volume} {66}} (\bibinfo
  {year} {2014})}\BibitemShut {NoStop}%
\bibitem [{\citenamefont {Phani}\ and\ \citenamefont
  {Hussein}(2017)}]{Phani_2017}%
  \BibitemOpen
  \bibfield  {author} {\bibinfo {author} {\bibfnamefont {A.~S.}\ \bibnamefont
  {Phani}}\ and\ \bibinfo {author} {\bibfnamefont {M.~I.}\ \bibnamefont
  {Hussein}},\ }\enquote {\bibinfo {title} {Introduction to lattice
  materials},}\ in\ \href {\doibase https://doi.org/10.1002/9781118729588.ch1}
  {\emph {\bibinfo {booktitle} {Dynamics of Lattice Materials}}}\ (\bibinfo
  {publisher} {John Wiley \& Sons, Ltd},\ \bibinfo {year} {2017})\
  Chap.~\bibinfo {chapter} {1}, pp.\ \bibinfo {pages} {1--17}\BibitemShut
  {NoStop}%
\bibitem [{\citenamefont {Barnes}\ \emph {et~al.}(2021)\citenamefont {Barnes},
  \citenamefont {Willey}, \citenamefont {Rosenberg}, \citenamefont {Medina},\
  and\ \citenamefont {Juhl}}]{Barnes_2021}%
  \BibitemOpen
  \bibfield  {author} {\bibinfo {author} {\bibfnamefont {C.~J.}\ \bibnamefont
  {Barnes}}, \bibinfo {author} {\bibfnamefont {C.~L.}\ \bibnamefont {Willey}},
  \bibinfo {author} {\bibfnamefont {K.}~\bibnamefont {Rosenberg}}, \bibinfo
  {author} {\bibfnamefont {A.}~\bibnamefont {Medina}}, \ and\ \bibinfo {author}
  {\bibfnamefont {A.~T.}\ \bibnamefont {Juhl}},\ }\bibfield  {title} {\enquote
  {\bibinfo {title} {Initial computational investigation toward passive
  transition delay using a phononic subsurface},}\ }in\ \href@noop {} {\emph
  {\bibinfo {booktitle} {AIAA Scitech 2021 Forum}}}\ (\bibinfo {year} {2021})\
  p.\ \bibinfo {pages} {1454}\BibitemShut {NoStop}%
\bibitem [{\citenamefont {Kianfar}\ and\ \citenamefont
  {Hussein}(2023)}]{kianfar2023phononicNJP}%
  \BibitemOpen
  \bibfield  {author} {\bibinfo {author} {\bibfnamefont {A.}~\bibnamefont
  {Kianfar}}\ and\ \bibinfo {author} {\bibfnamefont {M.~I.}\ \bibnamefont
  {Hussein}},\ }\bibfield  {title} {\enquote {\bibinfo {title}
  {Phononic-subsurface flow stabilization by subwavelength locally resonant
  metamaterials},}\ }\href@noop {} {\bibfield  {journal} {\bibinfo  {journal}
  {New Journal of Physics}\ }\textbf {\bibinfo {volume} {25}},\ \bibinfo
  {pages} {053021} (\bibinfo {year} {2023})}\BibitemShut {NoStop}%
\bibitem [{\citenamefont {Michelis}, \citenamefont {Putranto},\ and\
  \citenamefont {Kotsonis}(2023)}]{michelis2023attenuation}%
  \BibitemOpen
  \bibfield  {author} {\bibinfo {author} {\bibfnamefont {T.}~\bibnamefont
  {Michelis}}, \bibinfo {author} {\bibfnamefont {A.}~\bibnamefont {Putranto}},
  \ and\ \bibinfo {author} {\bibfnamefont {M.}~\bibnamefont {Kotsonis}},\
  }\bibfield  {title} {\enquote {\bibinfo {title} {Attenuation of
  tollmien--schlichting waves using resonating surface-embedded phononic
  crystals},}\ }\href@noop {} {\bibfield  {journal} {\bibinfo  {journal}
  {Physics of Fluids}\ }\textbf {\bibinfo {volume} {35}},\ \bibinfo {pages}
  {044101} (\bibinfo {year} {2023})}\BibitemShut {NoStop}%
\bibitem [{\citenamefont {Orr}(1907)}]{Orr1907}%
  \BibitemOpen
  \bibfield  {author} {\bibinfo {author} {\bibfnamefont {W.~M.}\ \bibnamefont
  {Orr}},\ }\bibfield  {title} {\enquote {\bibinfo {title} {The stability or
  instability of the steady motions of a perfect liquid and of a viscous
  liquid. part ii: A viscous liquid},}\ }in\ \href@noop {} {\emph {\bibinfo
  {booktitle} {Proceedings of the Royal Irish Academy. Section A: Mathematical
  and Physical Sciences}}},\ Vol.~\bibinfo {volume} {27}\ (\bibinfo
  {organization} {JSTOR},\ \bibinfo {year} {1907})\ pp.\ \bibinfo {pages}
  {69--138}\BibitemShut {NoStop}%
\bibitem [{\citenamefont {Sommerfeld}(1909)}]{Sommerfeld1909}%
  \BibitemOpen
  \bibfield  {author} {\bibinfo {author} {\bibfnamefont {A.}~\bibnamefont
  {Sommerfeld}},\ }\href@noop {} {\emph {\bibinfo {title} {Ein beitrag zur
  hydrodynamischen erklaerung der turbulenten fluessigkeitsbewegungen}}}\
  (\bibinfo {year} {1909})\BibitemShut {NoStop}%
\bibitem [{\citenamefont {Danabasoglu}, \citenamefont {Biringen},\ and\
  \citenamefont {Streett}(1991)}]{Dana91}%
  \BibitemOpen
  \bibfield  {author} {\bibinfo {author} {\bibfnamefont {G.}~\bibnamefont
  {Danabasoglu}}, \bibinfo {author} {\bibfnamefont {S.}~\bibnamefont
  {Biringen}}, \ and\ \bibinfo {author} {\bibfnamefont {C.~L.}\ \bibnamefont
  {Streett}},\ }\bibfield  {title} {\enquote {\bibinfo {title} {Spatial
  simulation of instability control by periodic suction blowing},}\ }\href
  {\doibase 10.1063/1.857896} {\bibfield  {journal} {\bibinfo  {journal}
  {Physics of Fluids A: Fluid Dynamics}\ }\textbf {\bibinfo {volume} {3}},\
  \bibinfo {pages} {2138--2147} (\bibinfo {year} {1991})}\BibitemShut {NoStop}%
\bibitem [{\citenamefont {Saiki}\ \emph {et~al.}(1993)\citenamefont {Saiki},
  \citenamefont {Biringen}, \citenamefont {Danabasoglu},\ and\ \citenamefont
  {Streett}}]{Saiki93}%
  \BibitemOpen
  \bibfield  {author} {\bibinfo {author} {\bibfnamefont {E.~M.}\ \bibnamefont
  {Saiki}}, \bibinfo {author} {\bibfnamefont {S.}~\bibnamefont {Biringen}},
  \bibinfo {author} {\bibfnamefont {G.}~\bibnamefont {Danabasoglu}}, \ and\
  \bibinfo {author} {\bibfnamefont {C.~L.}\ \bibnamefont {Streett}},\
  }\bibfield  {title} {\enquote {\bibinfo {title} {Spatial simulation of
  secondary instability in plane channel flow: comparison of {K}- and {H}-type
  disturbances},}\ }\href {\doibase 10.1017/S0022112093001879} {\bibfield
  {journal} {\bibinfo  {journal} {Journal of Fluid Mechanics}\ }\textbf
  {\bibinfo {volume} {253}},\ \bibinfo {pages} {485–507} (\bibinfo {year}
  {1993})}\BibitemShut {NoStop}%
\bibitem [{\citenamefont {Kucala}\ and\ \citenamefont
  {Biringen}(2014)}]{Kucala14}%
  \BibitemOpen
  \bibfield  {author} {\bibinfo {author} {\bibfnamefont {A.}~\bibnamefont
  {Kucala}}\ and\ \bibinfo {author} {\bibfnamefont {S.}~\bibnamefont
  {Biringen}},\ }\bibfield  {title} {\enquote {\bibinfo {title} {Spatial
  simulation of channel flow instability and control},}\ }\href {\doibase
  10.1017/jfm.2013.532} {\bibfield  {journal} {\bibinfo  {journal} {Journal of
  Fluid Mechanics}\ }\textbf {\bibinfo {volume} {738}},\ \bibinfo {pages}
  {105–123} (\bibinfo {year} {2014})}\BibitemShut {NoStop}%
\bibitem [{\citenamefont {Lighthill}(1958)}]{Lighthill_1958}%
  \BibitemOpen
  \bibfield  {author} {\bibinfo {author} {\bibfnamefont {M.~J.}\ \bibnamefont
  {Lighthill}},\ }\bibfield  {title} {\enquote {\bibinfo {title} {On
  displacement thickness},}\ }\href@noop {} {\bibfield  {journal} {\bibinfo
  {journal} {Journal of Fluid Mechanics}\ }\textbf {\bibinfo {volume} {4}},\
  \bibinfo {pages} {383--392} (\bibinfo {year} {1958})}\BibitemShut {NoStop}%
\bibitem [{\citenamefont {Sankar}, \citenamefont {Malone},\ and\ \citenamefont
  {Tassa}(1981)}]{Sankar_1981}%
  \BibitemOpen
  \bibfield  {author} {\bibinfo {author} {\bibfnamefont {N.~L.}\ \bibnamefont
  {Sankar}}, \bibinfo {author} {\bibfnamefont {J.~B.}\ \bibnamefont {Malone}},
  \ and\ \bibinfo {author} {\bibfnamefont {Y.}~\bibnamefont {Tassa}},\
  }\bibfield  {title} {\enquote {\bibinfo {title} {An implicit conservative
  algorithm for steady and unsteady three-dimensional transonic potential
  flows},}\ }in\ \href@noop {} {\emph {\bibinfo {booktitle} {AIAA Paper
  81-1016, June 1981}}}\ (\bibinfo {year} {1981})\BibitemShut {NoStop}%
\bibitem [{\citenamefont {Hussein}, \citenamefont {Hulbert},\ and\
  \citenamefont {Scott}(2006)}]{HusseinJSV06}%
  \BibitemOpen
  \bibfield  {author} {\bibinfo {author} {\bibfnamefont {M.~I.}\ \bibnamefont
  {Hussein}}, \bibinfo {author} {\bibfnamefont {G.~M.}\ \bibnamefont
  {Hulbert}}, \ and\ \bibinfo {author} {\bibfnamefont {R.~A.}\ \bibnamefont
  {Scott}},\ }\bibfield  {title} {\enquote {\bibinfo {title} {Dispersive
  elastodynamics of 1d banded materials and structures: analysis},}\
  }\href@noop {} {\bibfield  {journal} {\bibinfo  {journal} {Journal of Sound
  and Vibration}\ }\textbf {\bibinfo {volume} {289}},\ \bibinfo {pages}
  {779--806} (\bibinfo {year} {2006})}\BibitemShut {NoStop}%
\bibitem [{\citenamefont {Newmark}(1959)}]{Newmark}%
  \BibitemOpen
  \bibfield  {author} {\bibinfo {author} {\bibfnamefont {N.~M.}\ \bibnamefont
  {Newmark}},\ }\bibfield  {title} {\enquote {\bibinfo {title} {A method of
  computation for structural dynamics},}\ }\href@noop {} {\bibfield  {journal}
  {\bibinfo  {journal} {Journal of the Engineering Mechanics Division}\
  }\textbf {\bibinfo {volume} {85}},\ \bibinfo {pages} {67--94} (\bibinfo
  {year} {1959})}\BibitemShut {NoStop}%
\bibitem [{\citenamefont {Farhat}\ and\ \citenamefont
  {Lesoinne}(2000)}]{Farhat_2000}%
  \BibitemOpen
  \bibfield  {author} {\bibinfo {author} {\bibfnamefont {C.}~\bibnamefont
  {Farhat}}\ and\ \bibinfo {author} {\bibfnamefont {M.}~\bibnamefont
  {Lesoinne}},\ }\bibfield  {title} {\enquote {\bibinfo {title} {Two effcient
  staggered algorithms for the serial and parallel solution of
  three-dimensional nonlinear transient aeroelastic problems},}\ }\href@noop {}
  {\bibfield  {journal} {\bibinfo  {journal} {Computer Methods in Applied
  Mechanics and Engineering}\ }\textbf {\bibinfo {volume} {182}},\ \bibinfo
  {pages} {499--515} (\bibinfo {year} {2000})}\BibitemShut {NoStop}%
\bibitem [{\citenamefont {Prandtl}(1921)}]{Prandtl1921}%
  \BibitemOpen
  \bibfield  {author} {\bibinfo {author} {\bibfnamefont {L.}~\bibnamefont
  {Prandtl}},\ }\bibfield  {title} {\enquote {\bibinfo {title} {Bemerkungen
  {\"u}ber die entstehung der turbulenz},}\ }\href@noop {} {\bibfield
  {journal} {\bibinfo  {journal} {ZAMM-Journal of Applied Mathematics and
  Mechanics/Zeitschrift f{\"u}r Angewandte Mathematik und Mechanik}\ }\textbf
  {\bibinfo {volume} {1}},\ \bibinfo {pages} {431--436} (\bibinfo {year}
  {1921})}\BibitemShut {NoStop}%
\bibitem [{\citenamefont {Morris}(1976)}]{Morris_1976}%
  \BibitemOpen
  \bibfield  {author} {\bibinfo {author} {\bibfnamefont {P.~J.}\ \bibnamefont
  {Morris}},\ }\bibfield  {title} {\enquote {\bibinfo {title} {The spatial
  viscous instability of axisymmetric jets},}\ }\href@noop {} {\bibfield
  {journal} {\bibinfo  {journal} {Journal of Fluid Mechanics}\ }\textbf
  {\bibinfo {volume} {77}},\ \bibinfo {pages} {511--529} (\bibinfo {year}
  {1976})}\BibitemShut {NoStop}%
\bibitem [{\citenamefont {Morkovin}(1990)}]{morkovin1990roughness}%
  \BibitemOpen
  \bibfield  {author} {\bibinfo {author} {\bibfnamefont {M.~V.}\ \bibnamefont
  {Morkovin}},\ }\bibfield  {title} {\enquote {\bibinfo {title} {On
  roughness—induced transition: facts, views, and speculations},}\ }in\
  \href@noop {} {\emph {\bibinfo {booktitle} {Instability and Transition:
  Materials of the workshop held May 15-June 9, 1989 in Hampton, Virgina Volume
  1}}}\ (\bibinfo {organization} {Springer},\ \bibinfo {year} {1990})\ pp.\
  \bibinfo {pages} {281--295}\BibitemShut {NoStop}%
\bibitem [{\citenamefont {Cossu}\ and\ \citenamefont
  {Brandt}(2004)}]{Cossu_2004}%
  \BibitemOpen
  \bibfield  {author} {\bibinfo {author} {\bibfnamefont {C.}~\bibnamefont
  {Cossu}}\ and\ \bibinfo {author} {\bibfnamefont {L.}~\bibnamefont {Brandt}},\
  }\bibfield  {title} {\enquote {\bibinfo {title} {On
  {T}ollmien–{S}chlichting-like waves in streaky boundary layers},}\
  }\href@noop {} {\bibfield  {journal} {\bibinfo  {journal} {European Journal
  of Mechanics B/Fluids}\ }\textbf {\bibinfo {volume} {23}},\ \bibinfo {pages}
  {815--833} (\bibinfo {year} {2004})}\BibitemShut {NoStop}%
\end{thebibliography}%

\appendix
\renewcommand\theequation{A\arabic{equation}}
\renewcommand\thefigure{A\arabic{figure}}
\renewcommand\thesection{A\arabic{section}}  

\setcounter{figure}{0}      
\setcounter{equation}{0}  
\setcounter{section}{0}  

\section{Flow-PSub modeling information}
\label{sec:appendix1}
The flow simulations we run in this investigation are based on the fully nonlinear three-dimensional Navier-Stokes equations for incompressible channel flows, which are 
\begin{equation}
\label{eq:cont}
\frac{\partial u_i}{\partial x_i} = 0
\end{equation}
for continuity, and
\begin{equation}
\label{eq:NS}
\frac{\partial u_i}{\partial t} + \frac{\partial u_j u_i}{\partial x_j} = \frac{2}{Re}-\frac{\partial p}{\partial x_i} + \frac{1}{Re}\frac{\partial u_i}{\partial x_j x_j}
\end{equation}
for momentum conservation, respectively. Within the control region, the coupling FPI boundary conditions  are specified as
\begin{subequations}
\label{eq:structbc}
	\begin{equation}
	\label{eq:structbcu}
	u(x_{\rm s}\le x\le x_{\rm e},y=0,z,t)=-\frac{\eta(s=0,t^*)}{\delta}\frac{\mathrm{d}u_{\rm b}}{\mathrm{d}y}, 
	\end{equation}
	\begin{equation}
	\label{eq:structbcv}
	v(x_{\rm s}\le x\le x_{\rm e},y=0,z,t)=\frac{\dot{\eta}(s=0,t^*)}{U_\mathrm{c}}.
	\end{equation}
\end{subequations}
These are imposed to ensure the stresses and velocities match at the interface.\cite{Hussein_2015}~In Eq.~\eqref{eq:structbc}, $\eta(s, t^*)$ and $\dot{\eta}(s, t^*)$ stand for the displacement and velocity of the PSub, respectively, where $s$ is the structure's axial spatial coordinate and $t^*$ is (dimensional) time.~Referred to as transpiration boundary conditions,~\cite{Lighthill_1958,Sankar_1981} Eqs.~(\ref{eq:structbcu}) and~(\ref{eq:structbcv}) are obtained by keeping the interface location fixed and retaining only the linear terms following a Taylor series expansion of the exact interface compatibility conditions. Equations~\eqref{eq:structbc} are valid if the PSub motion is only in the wall-normal direction and $\eta \ll \delta$. Hence, throughout DNS the roughness Reynolds number is monitored and maintained below 25.\cite{morkovin1990roughness} \\
\indent The PSub axial displacement, velocity, and acceleration are obtained by solving the governing equation for a one-dimensional linear elastic slender rod structure
\begin{equation}
    \rho_{\rm s} \ddot{\eta}=(E\eta_{,s}+C\dot{\eta}_{,s})_{,s}+f,
    \label{eq:structure}
\end{equation}
 where $\rho_{\rm s}=\rho_{\rm s}(s)$, $E=E(s)$, and $C=C(s)$, respectively, represent density, elastic modulus, and damping of the PSub. In Eq.~\eqref{eq:structure}, $f$ is the external forcing on the PSub, either artificially applied in the pre-simulation PSub characterization analysis or imposed by the dimensional spatial-average flow pressure within the control region during the coupled fluid-structure simulations. In Eq.~\eqref{eq:structure}, the differentiation with respect to position is indicated by $(.)_{,s}$, and the superposed single dot $\dot{(.)}$ and double dot $\ddot{(.)}$ denote the first and second-time derivatives, respectively. Free-fixed-end boundary conditions are applied on the PSub, while the forcing is applied to the free (top) end at $s=0$.\\ 
 \indent During the coupled fluid-structure simulations, the pressure field acting on the surface over a given PSub is integrated at each time step to yield a value for $f$. This forcing value is applied as an input to the PSub solver to produce the responding velocity at $s=0$. This velocity is then applied to the flow uniformly across the region where the individual PSub interfaces with the flow.

\section{Flow and PSub quantities obtained by post-processing the simulations data}
\label{sec:appendix2}
Throughout the investigation, several quantities of interest are calculated by post-processing the time-dependent data emerging from the coupled fluid-structure simulations. \\
\indent On the flow side, we calculate the streamwise position-dependent integral of the perturbation kinetic energy, which is defined as
\begin{equation}
K_\mathrm{p}\left(x\right)= \int_0^{2\pi} \int_0^{2} \frac{1}{2}\left(\left\langle\hat{u}^ 2\right\rangle+\left\langle\hat{v}^{2}\right\rangle+\left\langle\hat{w}^{2}\right\rangle\right) \mathrm{d} y \mathrm{~d} z,
\label{eq:Kp}
\end{equation}
where $\left(\langle \cdot \rangle\right)$ and $\left(\hat{\cdot}\right)$ represent the time-averaged and perturbation quantities, respectively. Alternatively, we calculate the $x-y$ contour of the perturbation kinetic energy as
\begin{equation}
k_\mathrm{p}\left(x,y\right)= \int_0^{2\pi} \frac{1}{2}\left(\left\langle\hat{u}^ 2\right\rangle+\left\langle\hat{v}^{2}\right\rangle+\left\langle\hat{w}^{2}\right\rangle\right) \mathrm{~d} z.
\label{eq:kp}
\end{equation}
\noindent We also calculate the rate of production of perturbation kinetic energy,~\cite{Prandtl1921,Cossu_2004} which in dimensionless form is 
 \begin{equation}
    P_{\rm r}(x,y) =  - \langle \hat{u} \hat{v}\rangle \frac{\partial \langle u \rangle}{\partial y}.
\end{equation}
Along the surface, we are interested in the wall shear stress and friction drag. The time-averaged wall shear stress is defined as
\begin{equation}
\langle \tau_w (x) \rangle = \left[\mu_f \frac{\partial \langle u \rangle}{\partial y} - \rho_f \langle \hat{u} \hat{v} \rangle\right]_{y=0}, 
\end{equation}
in which on the right hand side the first term represents the mean shear stress, and the second term represents the perturbation (Reynolds) shear stress at the wall. Note, the second term is naturally zero for the rigid wall, whereas within the control region, even though substantially smaller than the mean shear stress, it is non-zero due to the PSub's elastic deformation.
Also, the streamwise position-dependent friction drag is obtained by integrating the skin friction as follows

\begin{equation}
    D(x) = \frac{1}{2}\rho_f U_b \int_0^x  C_f(\xi) \mathrm{d}\xi,
\end{equation}
where $U_b$ is the bulk velocity and $C_f(x)={2 \langle \tau_\mathrm{w}(x) \rangle}/{\rho_f U_b}$ is the skin friction coefficient. \\
\indent On the PSub side, we are interested in the elastodynamic energy within the PSub elastic domain, which is specified as
\begin{equation}
    \Psi(s,t)=\frac{1}{2}\left[E(\mathrm{d}\eta/ \mathrm{ds})^2+\rho_s \dot{\eta}^2\right].
\end{equation}
\begin{figure} [t!]
\centering
\includegraphics{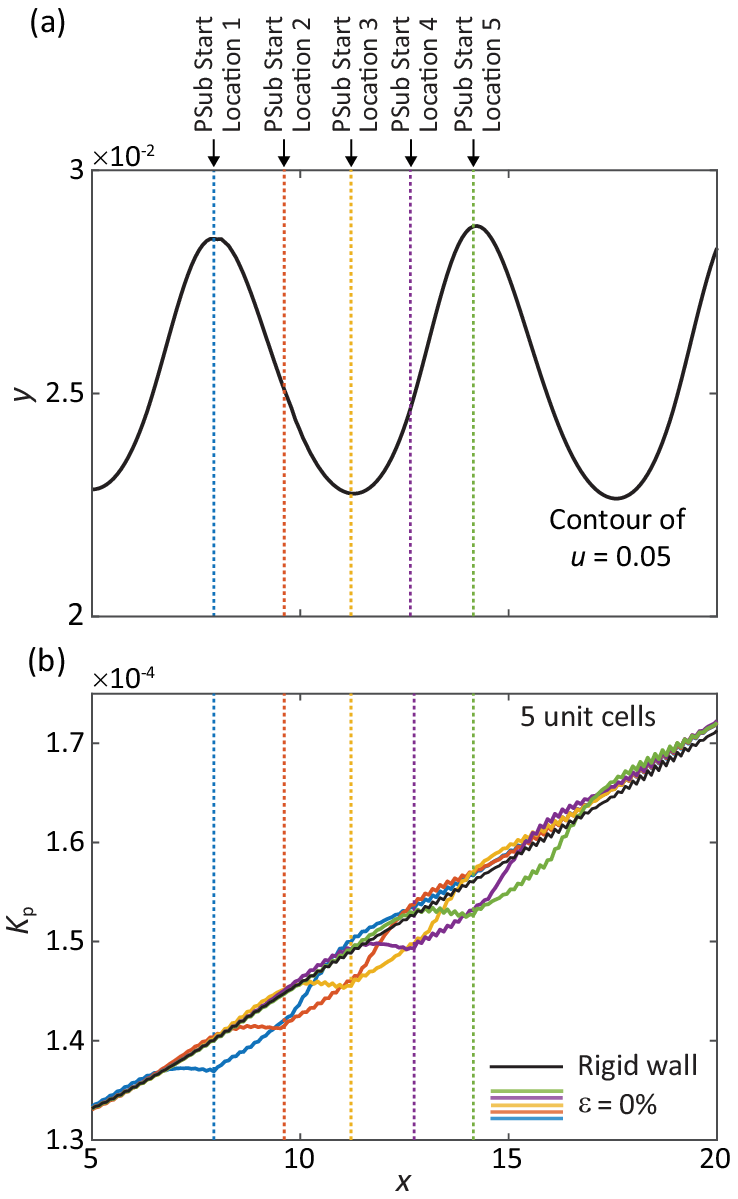}
\caption{\label{fig:figA1} Demonstration of responsiveness of a PSub regardless of its position with respect to the incoming instability wave. (a) Contour of the total (instantaneous) velocity $u=0.05$ line in the $x-y$ plane with markings indicating the streamwise position of $x = x_s$ for the PSub, considering five different locations labeled Start Location 1 to Start Location 5. A separate simulation is executed for each of these five cases. (b) Streamwise spatial distribution of $K_\mathrm{p}$ for each PSub location.~These results are for the $\varepsilon=0\%$ (moderate stabilization) PSub design.}
\end{figure}
\section{Responsiveness of PSubs}
\label{sec:appendix3}
Figure \ref{fig:figA1} explores the behavior of PSubs as passively \textit{responsive} subsurface structures that provide control irrespective of the incoming phase. Here, we have installed a single PSub at several streamwise $x$ locations, each representing an independent case. Figure \ref{fig:figA1}(a) shows the leading edge of any given PSub, i.e., the $x=x_s$ position, relative to the streamwise position of the total (instantaneous) velocity contour $u=0.05$. The wavy motion of the flow field can be seen in Fig. \ref{fig:figA1}(a) due to the introduced spatially unstable TS wave at the inlet of the channel. We observe that as $x=x_s$ (for different PSub locations considered in separate simulations) traverses downstream, the wall-normal integrated perturbation kinetic energy $K_\mathrm{p}$ as a function of $x$ behaves qualitatively the same. Quantitatively, there is a slight enhancement in the local reduction of $K_\mathrm{p}$ for the PSubs installed downstream, merely because the amplitude of the TS waves (slowly) grows spatially, hence $P(\Omega_{\mathrm{TS}})$ becomes higher resulting in a greater PSub influence. However, the effectiveness of the PSub control is fundamentally independent of its relative position with respect to the $u$ value of the TS wave. In fact, a PSub passively adjusts its motion to be always out-of-phase (or in-phase) of the existing disturbances for stabilization (or destabilization). This phenomenon underlines the robust responsiveness and adaptivity of the PSubs independent of where they are installed with respect to the instantaneous waveform of the instability.
\begin{figure} [b!]
\centering
\includegraphics{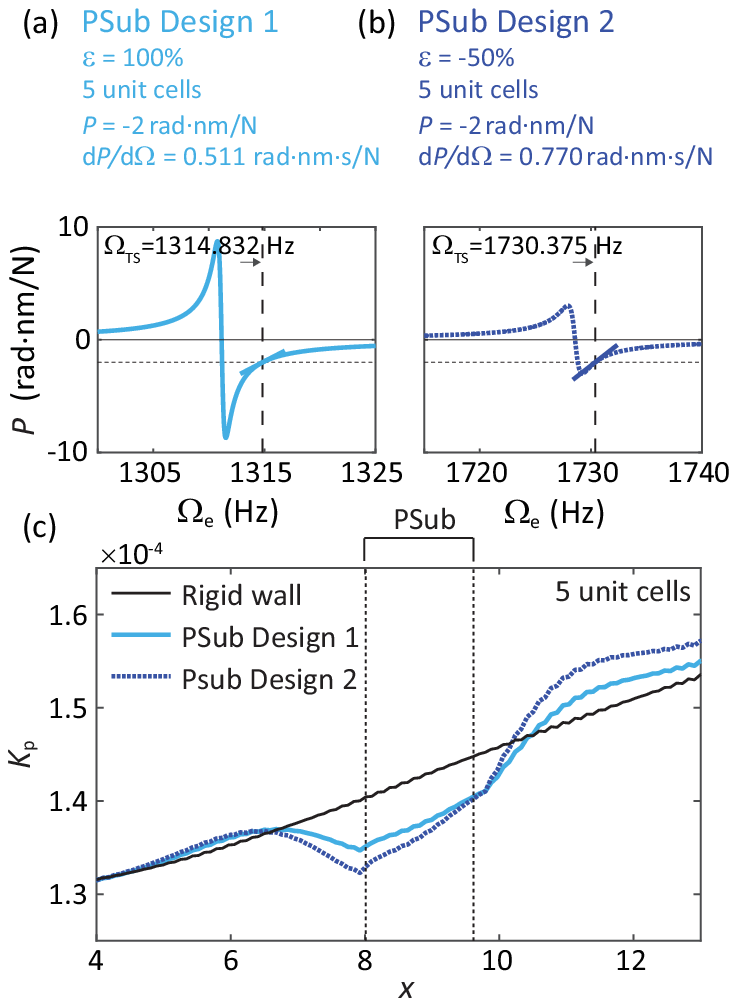}
\caption{\label{fig:figA2} Performance metric analysis. Evaluation of $P$ and $\mathrm{d}P/\mathrm{d}\Omega$ for two PSub designs: (a) PSub Design 1, and (b) PSub Design 2. The TS wave frequencies are selected such that the value of $P$ matches exactly in both designs. (c) Flow perturbation kinetic energy $K_\mathrm{p}$ as a function of streamwise position $x$ for two PSub designs operating at the corresponding selected TS wave frequencies. The all-rigid-wall case is also provided for comparison.}
\end{figure}
\section{Effect of PSub performance metric: $P$ versus $dP / d\Omega$}
\label{sec:appendix4}
Hussein et al.~\cite{Hussein_2015} have clearly shown that there is a direct relationship between the value and sign of the performance metric $P$ and how a PSub responds to the flow disturbances at any given frequency $\Omega_{\mathrm{ST}}$. Recall, a positive $P$ value corresponds to an in-phase response causing destabilization, whereas a negative $P$ value means the PSub's motion is out-of-phase with respect to the flow disturbances which enables stabilization. Here we consider also the correlation between the frequency derivative of the performance metric $\mathrm{d}P/\mathrm{d}\Omega$ and the flow instability response.  To distinguish between the correlation with $P$ versus $\mathrm{d}P/\mathrm{d}\Omega$, we consider  two different PSub designs. PSub Design 1 targets a TS wave of frequency $\Omega_\mathrm{TS}=1314.832$ $\mathrm{Hz}$ and PSub Design 2 targets a TS wave of frequency $\Omega_\mathrm{TS}=1730.375$ $\mathrm{Hz}$. These two designs are chosen to give the same value of the performance metric $P(\Omega_{\mathrm{TS}})=-2$ rad$\cdot$nm/N, but slightly different corresponding $\mathrm{d}P/\mathrm{d}\Omega$ values, as illustrated in Figs.~\ref{fig:figA2}(a) and \ref{fig:figA2}(b). 
According to Fig.~\ref{fig:figA2}(c), at the trailing edge of the PSub control region, $x=x_e$, the value of $K_\mathrm{p}$ is roughly the same for both PSub designs. However, and importantly, PSub Design 2$-$with a higher $\mathrm{d}P/\mathrm{d}\Omega$ value$-$causes a stronger reduction in $K_\mathrm{p}$ within the control region, mostly at $x=x_s$. This is despite the two designs exhibiting exactly the same value of $P$. This shows that the $\mathrm{d}P/\mathrm{d}\Omega$ metric provides an additional degree of predictive power on the actual response in the coupled flow-PSub simulations. Consistent with the results of Fig.~\ref{fig:fig02}(i), a greater value of $\mathrm{d}P/\mathrm{d}\Omega$ predicts a higher perturbation growth rate within the control region. Moreover, downstream of the control region, it is shown that for the same $P$, $K_\mathrm{p}$ increases further with an increase in $\mathrm{d}P/\mathrm{d}\Omega$. 
\begin{figure}[b!]
\centering
\includegraphics{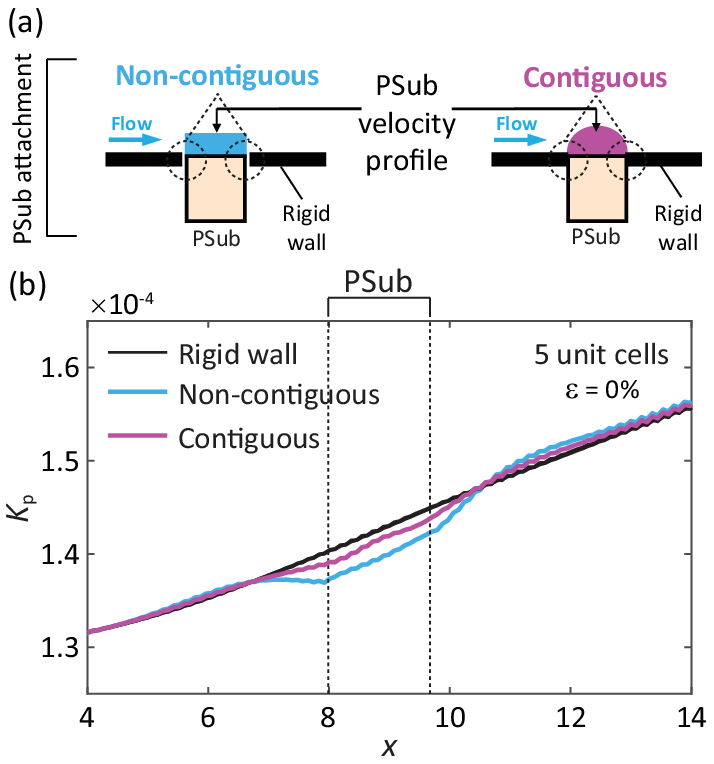}
\caption{\label{fig:figA3} Non-contiguous versus contiguous PSub edge attachment conditions. (a) Schematic of each attachment approach. In the non-contiguous attachment, the PSub motion is independent of the rigid wall. In the contiguous attachment, the PSub motion at the edges is zero and varies smoothly toward the center. (b) Flow instability kinetic energy $K_\mathrm{p}$ as a function of streamwise position $x$ for the two attachment approaches.}
\end{figure}
\section{Effect of PSub type of connection to rigid wall}
\label{sec:appendix5}
In most of the results provided, a uniform velocity field (obtained from FE analysis) is assigned to all nodes along the fluid-PSub interface for each PSub at each time step. Recall, the integrated pressure within the control region is fed to the PSub model at each time step, and the PSub response (displacement, velocity, and acceleration) is returned to the flow field in the fluid model by applying the transpiration boundary conditions, Eqs.~\eqref{eq:structbcu} and \eqref{eq:structbcv}. In our default approach, all the computational grid points along the control region are fed the same velocity value for a given PSub. This creates a discontinuity in the velocity profile at the edges of the PSub, where the PSub meets the rigid wall or an adjacent PSub. We refer to this as a non-contiguous connection. Here we explore an alternative connection, where we assume the PSub is contiguously attached to the rigid wall. This is implemented by introducing a smooth polynomial fit across the control region for each PSub, having the PSub FE response prescribed to the grid point at the center of the PSub edge. This fit enforces zero values with zero slopes at the start and end edges where the PSub is connected to the rigid wall. With these constraints, a polynomial of order four is assigned within the PSub control region yielding a smooth transition to the rigid wall at the PSub edges. A comparison between this contiguous and the default non-contiguous connections is demonstrated in Fig.~\ref{fig:figA3}. As expected, the contiguous connection yields a smoother profile for the instability kinetic energy. We also note that the profile experiences a maximum at nearly the middle of the PSub domain. Most important is the significantly improved recovery downstream of the PSub, which is advantageous for the use of multiple PSubs to extend the spatial domain of control. Future research will examine two-dimensional or three-dimensional PSub FE models, which would eliminate the need for polynomial fitting for contiguous connections.    

\end{document}